\documentclass[twocolumn]{emulateapj}
\lefthead{Peter et al. 2007}
\righthead{Morphologies of Protocluster Galaxies}

\newcommand{\minpoint}{\mbox{$'\mskip-4.7mu.\mskip0.8mu$}}
\newcommand{\secpoint}{\mbox{$''\mskip-7.6mu.\,$}}

\begin{document}

\title{Morphologies of Galaxies in and around a Protocluster at $z=2.300$}
\author{ Annika H. G. Peter}
\affil{ Princeton University, Jadwin Hall -- Washington Road, 
Department of Physics, Princeton, NJ, 08544}
\author{ Alice E. Shapley}
\affil{ Princeton University, Peyton Hall -- Ivy Lane, Department of 
Astrophysical Sciences, Princeton, NJ 08544}
\author{ David R. Law \& Charles C. Steidel}
\affil{California Institute of Technology, MS 105--24, Pasadena, CA 91125}
\author{ Dawn K. Erb}
\affil{Harvard-Smithsonian Center for Astrophysics, MS 20, 
60 Garden St., Cambridge, MA 02138}
\author{ Naveen A. Reddy}
\affil{National Optical Astronomy Observatory, 950 North Cherry Ave, 
Tuscon, AZ, 85719}
\author{ Max Pettini}
\affil{Institute of Astronomy, University of Cambridge, Madingley Road, 
Cambridge, CB3 0HA, UK}

\begin{abstract}
We present results from the first robust
investigation of galaxy morphology as a function of environment at $z>1.5$. Our study is motivated 
by the fact
that star-forming galaxies contained within a protocluster at $z=2.3$ in the HS1700+64 field are 
characterized by
significantly older ages and larger stellar masses on average
than those at similar redshifts but more typical
environmental densities. In the analysis of \emph{HST}/ACS images, we apply non-parametric
statistics to characterize the rest-frame UV morphologies
of a sample of 85 UV-selected star-forming
galaxies at $z=1.7-2.9$, 22 of which
are contained in the protocluster.  The remaining 63 galaxies are not in the protocluster but have 
a similar mean redshift of $ \bar{z}\sim 2.3$, and 
constitute our control sample.  We find
no environmental dependence for the distributions
of morphological properties. Combining the measured morphologies
with the results of population synthesis modeling, 
we find only weak correlations, if any, between morphological properties
and measures of stellar population 
such as stellar mass, age, extinction and star-formation
rate.  These findings are similar to results from other recent work. 
Given the incomplete census of the protocluster galaxy population,
which is weighted towards star-forming galaxies,
and the lack of correlation between rest-frame UV morphology
and star-formation history at $z\sim 2$ within our
sample, the absence of environmental trends in the distribution
of morphological properties is not surprising.
We do, however, find some non-environmental trends with morphology.
The degree of nebulosity and apparent rest-frame UV size appears to correlate
with the UV-luminosity (uncorrected for dust extinction),
which we demonstrate is not an artifact of finite signal-to-noise.
Additionally, using a larger sample of photometric candidates, most
of which do not have spectroscopic redshifts, we compare
morphological distributions for 282 UV-selected
and 43 near-IR-selected galaxies.  While the observed difference in the degree of nebulosity 
between the two samples appears to be a byproduct of the fainter average rest-frame UV surface 
brightness of the near-IR-selected galaxies, we find that, among the lowest surface brightness 
galaxies, the near-IR-selected objects have significantly smaller angular sizes.  
\end{abstract}


\section{Introduction}
Galaxies populate the local universe in a very regular way.  Dynamically hot elliptical galaxies 
are found preferentially in large galaxy clusters, especially the very centers 
\citep{dressler1980,postman1984,goto2003}.  Dynamically cold spiral galaxies tend to lie in the 
less dense environments of groups or the field.  The name given for these trends is the 
``morphology-density'' relation.  The morphology-density relation is already in place by $z=1$, 
although there are quantitative, but not qualitative, differences between $z=1$ and $z=0$ 
\citep{postman2005,smith2005}.  For example, the elliptical fraction in clusters is lower at $z=1$ 
than at $z=0$, but the cluster elliptical fraction is higher than the elliptical fraction in the 
field in both epochs.  One of the outstanding problems in the theory of galaxy evolution and 
structure formation is understanding when and how this relation is established. 
\newline\indent
The morphology-density relation is entangled with several other observed relations between physical 
properties of galaxies and their classification and environment.  In the local universe, elliptical 
galaxies are structurally different and, on average, have higher mass, less active star formation 
(as well as less gas available for star formation), and redder colors than spiral galaxies 
\citep{kauffmann2003}.  Therefore, in addition to a morphology-density relation, strong trends of 
galaxy color with environment are observed in the Sloan Digital Sky Survey 
\citep[SDSS;][]{kauffmann2004,blanton2005}. These trends
have been traced out to $z = 1.3$ with the DEEP2 galaxy redshift survey
\citep{davis2003,cooper2006a,cooper2006b,gerke2006}.  Studies of the local universe
strongly imply that the correlations between color and star 
formation history and environment are more fundamental than the morphology-density relation 
\citep{kauffmann2004,blanton2005}.  As part of the program of understanding galaxy evolution, it is 
important to explain the relation between the physical structure of galaxies and their star 
formation history, and the correlations between the various galactic properties and environment.    
\newline\indent
There are many theories to explain how such correlations arise as part of the hierarchical picture 
of structure formation.  Several authors have considered how much of the morphology-density 
relation is simply a result of the merger histories of the host halos.  \citet{benson2001} graft 
semi-analytic models onto a cosmological dark matter N-body simulation 
\citep{jenkins1998,kauffmann1999} and examine the resulting morphology-density and color-density 
relations at the current epoch.  They find qualitative agreement with observations of the 
early-type fraction and average galaxy color as a function of environment.  Their semi-analytic 
model overpredicts the red fraction at the center of clusters, and underpredicts the number of 
spheroids in the field at $z=1$ \citep{benson2002}.  \citet{okamoto2001} also use semi-analytic 
models to interpret their dark matter simulations, but focus only on the morphology-density 
relation within clusters.  While they can reproduce the elliptical fraction as a function of 
density with their semi-analytic prescription, the modeled S0 fraction does not agree with 
observations.  Using a simple prescription for the relationship between the host dark matter halo 
mass accretion history and galaxies lying in the halos, \citet{maulbetsch2007} are able to 
qualitatively, but not quantitatively, reproduce the morphology-density and specific star formation 
rate environmental correlation at $z=0$.  Additionally, the predicted morphology-density relation 
predicted at $z=1$ differs dramatically from what is observed at that redshift.  Therefore, while 
it is likely that the merger history influences the correlations among environment, star formation 
history, and morphology, it is not the only driver.
\newline\indent
In particular, a number of baryonic processes are invoked to explain these correlations.  Each     
process is associated with a characteristic timescale and halo mass, producing different effects
on the photometric and structural properties of either central or satellite galaxies.  
Central galaxies continue to form stars as long as they can accrete significant amounts of cold gas. 
Cosmological simulations suggest that gas can cool in halos of $\mathrm{M}
< 10^{12} \mathrm{M}_\odot$ up to the current epoch, but not in   
more massive halos since $z\sim 2$ \citep{keres2005,croton2006,dekel2006}. This trend
translates directly into the color distributions of the associated galaxies, but
there is no prediction for how the gas accretion history specifically affects morphology evolution.  
Satellite galaxies
falling into more massive halos are subject to a number of processes, such as ram pressure stripping
\citep{gunn1972,abadi1999,hester2006}, ``strangulation'' or ``starvation''
\citep{larson1980,bekki2002}, and
harassment due to high-speed encounters with other galaxies \citep{moore1996,moore1999}.  The
relative importance of these mechanisms for altering morphology depends on both properties of the
satellite galaxies and the larger host halo.
\newline\indent
Disentangling these processes is extremely difficult, since galaxies can experience most of 
them during some point in their evolution.  In order to determine which processes drive the 
morphology-density, color-density, and star formation-density relations, it is important to observe 
galaxies in a range of environments at a variety of epochs, in which different processes may 
dominate.  In particular, since many theories on galaxy evolution focus on processes that are most 
likely to be effective in cluster environments, where the environmental contrast is maximized, it 
is important to observe clusters in different stages of formation, even before the structures have 
virialized.
\newline\indent
Several high-redshift galaxy overdensities have been observed that are thought to virialize into 
massive clusters by the current epoch.  Commonly called ``protoclusters,'' some of these objects 
have been found serendipitously \citep{pascarelle1996,steidel1998,steidel2005}. Others were 
discovered through narrow-band Ly$\alpha$ and H$\alpha$ surveys targeting the regions about high 
redshift radio galaxies \citep[HzRGs,][]{kurk2002,overzier2006} because HzRGs lie in highly biased 
environments \citep{debreuck2000}.  One of the best-established (and most thoroughly observed) 
protoclusters is at $z=2.300$ in the Q1700 field centered on the HS1700+643 quasar, and was 
discovered in a survey of $z=2.4\pm 0.4$ star-forming galaxies in that field 
\citep{steidel2004,steidel2005}.  Spectroscopy of $> 100$ photometric $z>1.5$ candidates in the 
Q1700 field allows this structure to be well-mapped in projected and redshift-space, and leads to a 
good estimate of the mass overdensity.  This stands in contrast to work on the other protoclusters, 
for which the estimated mass overdensity is based on projected number densities of Ly$\alpha$ 
emitters, which do not fully trace the underlying population of high redshift galaxies, and for 
which the number density in average environments is not well quantified.  Furthermore, protocluster 
membership for galaxies that are not Ly$\alpha$ emitters is established using photometric 
redshifts, which have large $\delta(z) \ge 0.5$ errors \citep{overzier2006}.  The mass overdensity 
of the Q1700 protocluster indicates that it should virialize into a massive cluster of $\sim 1.4 
\times 10^{15} \mathrm{M}_\odot$ by the present \citep{steidel2005}.  
\newline\indent
Stellar population synthesis modeling of the deep multiband photometry and spectroscopy of the 
star-forming galaxies in the Q1700 field indicate that galaxies in the protocluster are, on 
average, significantly more massive and older than galaxies in the field.  Complementary analyses 
of the spatial clustering of $z > 2$ galaxies as a function of rest-frame optical color suggest 
that redder galaxies are more strongly clustered \citep{adelberger2005b,quadri2007}, which is 
consistent with the results from the Q1700 protocluster.  Therefore, it appears that star-formation 
history is already a strong function of environment at high redshift. 
\newline\indent
In this paper, we explore connections between morphology and environment, and morphology and star 
formation history at high redshift.  We examine the rest-UV morphologies of optically-selected, 
spectroscopically confirmed star-forming galaxies in the Q1700 field, performing the first rigorous 
analysis of morphology as a function of environment at $z > 1.3$.  In addition, given the large 
multiband data set available for this field, we are able to identify a complementary population of 
high redshift galaxies selected by their near-infrared colors, and compare their properties with 
those of the UV-selected galaxies.  The structure of the paper is as follows: in Section 2, we 
summarize previous observations of the Q1700 field and present our new \emph{Hubble Space 
Telescope} Advanced Camera for Surveys (\emph{HST}/ACS) observations, from which we derive 
morphologies for our galaxy sample.  In Section 3, we describe the set of parameters we use to 
quantify morphology, and present tests of the robustness of these parameters.  We discuss the 
results of the analysis on the environmental dependence on morphology for star-forming galaxies in 
Section 4.  We present an analysis of morphology as a function of star formation history and the 
physical properties of galaxies in Section 5.  In Section 6, we compare the distributions of 
morphological parameters of optically and near-infrared selected galaxies.   We discuss the results 
of our analyses and place them in the broader picture of structure formation and galaxy evolution 
in Section 7.
\newline\indent
For the rest of this paper, we assume a flat cosmology with $\Omega_m = 0.3, \Omega_\Lambda = 0.7$, 
and $h = 0.7$.

\section{Observations and Sample Selection}
\subsection{Prior Observations}\label{sec:obs}
In order to optically select high redshift star-forming galaxies in the foreground of the 
HS1700+643 quasar, \citet{steidel2004} observed a $15\minpoint3 \times 15\minpoint3$ field (Q1700) 
with the Prime Focus Imager on the William Herschel Telescope in the $U_n$, G, and $\mathcal{R}$ 
bands, with supplementary observations using the $U_n$ filter with the Low Resolution Imaging 
Spectrometer (LRIS) on the Keck I telescope.   
\newline\indent
%Spectroscopic follow-up was concentrated on the region nearest the HS1700+643 quasar.  As presented in \citet{steidel2005}, spectroscopic redshifts were found for 100 photometric candidates using the LRIS-B instrument on Keck I.  In addition, \citet{erb2006b} determined redshifts for more objects using NIRSPEC near-infrared spectrograph on Keck II.  In total, 167 photometric galaxy candidates have secure redshifts, of which 141 have $z>1.4$.  The contaminants were stars and low-redshift galaxies.
Spectroscopic follow-up was concentrated on the region nearest the HS1700+643
quasar. Redshifts were primarily measured from optical spectra probing
the rest-frame UV. These spectra were
obtained using the LRIS-B instrument on the Keck~I telescope
\citep{steidel2004}. In addition,
\citet{erb2006b}  measured
H$\alpha$ redshifts for 19 objects in the larger optical spectroscopic
sample using the NIRSPEC near-infrared spectrograph on Keck~II.
In total, 167 galaxy photometric candidates have secure redshifts, 
of which 141 have $z>1.4$. The contaminants were stars and low-redshift
galaxies. The current Q1700 spectroscopic
sample is larger than the original one presented
in \citet{steidel2005}, due to subsequent additional observations.
\newline\indent
In order to perform population synthesis modeling on those galaxies with secure redshifts, near- 
and mid-infrared observations of the Q1700 field were performed and presented by 
\citet{barmby2004}, \citet{shapley2005a} and \citet{erb2006b}.  Data in the $3.5-8.0 \mu$m 
wavebands were obtained with the Infrared Array Camera (IRAC) on the \emph{Spitzer Space Telescope} 
during the In-Orbit Checkout (IOC).  Data reductions and photometry are described in 
\citet{barmby2004} and \citet{shapley2005a}.  Using the Wide Field Imaging Camera (WIRC) on the 5m 
Palomar Hale telescope, a $8\minpoint5 \times 8\minpoint5$ section of the Q1700 field was imaged in 
the J and $K_s$ \citep{erb2006b,shapley2005a} bands.  These data also allow us to identify a sample 
of distant red galaxies (``DRGs'') satisfying the $J-K_s > 2.3$ criterion of \citet{franx2003}.  
This selection criterion is tuned to select galaxies with significant 4000$\mbox{\AA}$/Balmer 
breaks at $z\approx 2.5$, and hence, selects galaxies with either old or heavily obscured stellar 
populations \citep{fs2004,vandokkum2004,reddy2005}.
%\newline\indent
%The Q1700 field was also imaged with the Large Field Camera on the Palomar Hale telescope using a narrowband filter (centered at 4010 $\mbox{\AA}$) tuned to the Lyman $\alpha$ transition at z=2.3 \citep[][, private communication]{bogosaljevic2006}.  These observations have allowed us to identify a sample of candidate Lyman$-\alpha$ emitters.

%%% Observation footprint
\begin{figure*}
%\figurenum{fig:foot}
\plotone{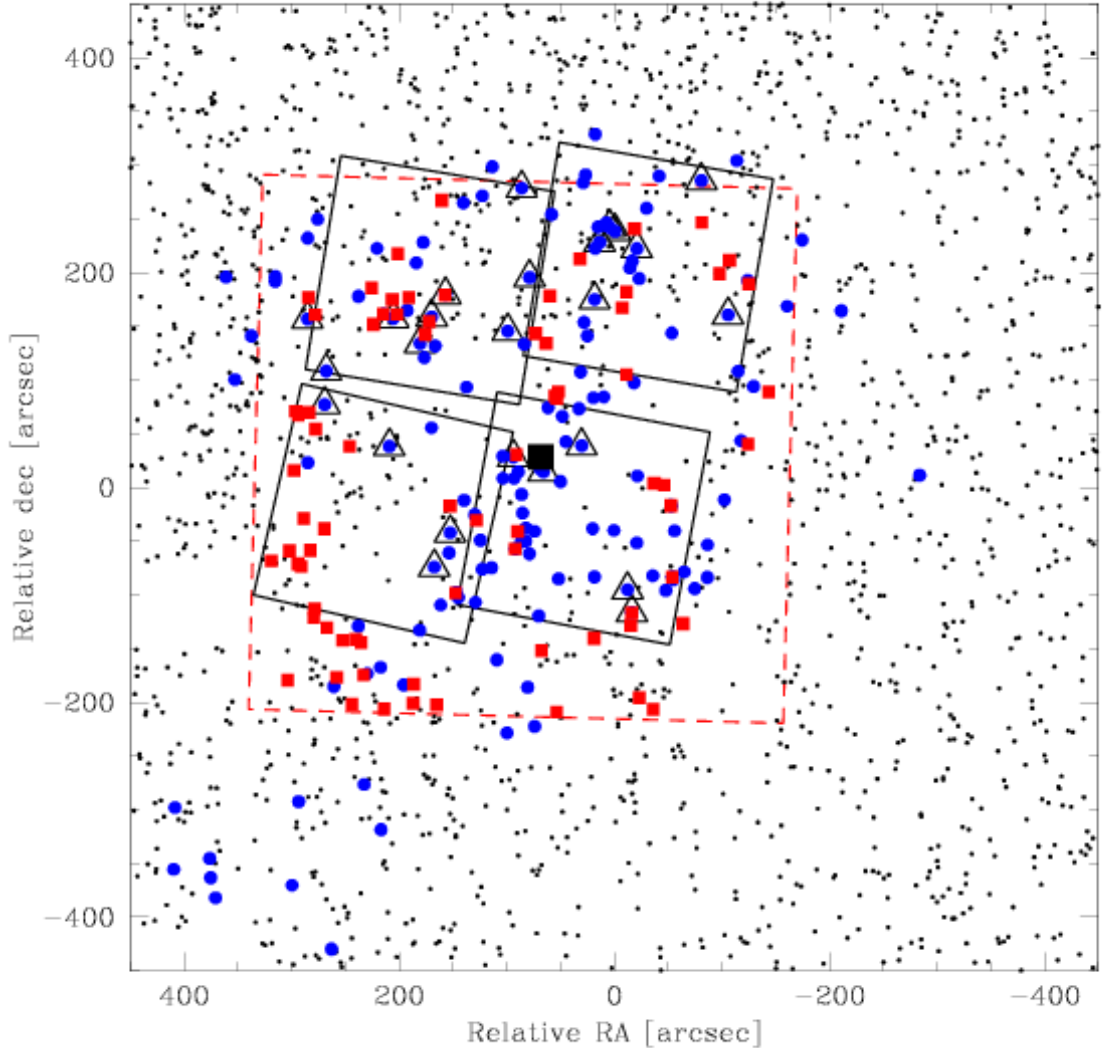}
\caption{Schematic of observations and samples.  The image covers the full scale of the $\mathcal{R}$-band observation.  The $K_s$ observation is bounded by the red box.  The \emph{HST}/ACS footprint is denoted by the four black-rimmed boxes.  Marked by the large solid black square is the HS1700+643 quasar.  Objects that satisfy the rest-frame UV color criteria but for which no redshift information is available are marked by small black dots; those with redshift information are denoted by large blue dots, with triangles about the spike objects.  DRGs are marked with red squares.  Objects satisfying more than one selection criterion will have stacked markers.}
\label{fig:foot}
\end{figure*}

%\clearpage
%%% Spike thumbnails
\begin{figure*}
\epsscale{1.15}
\plotone{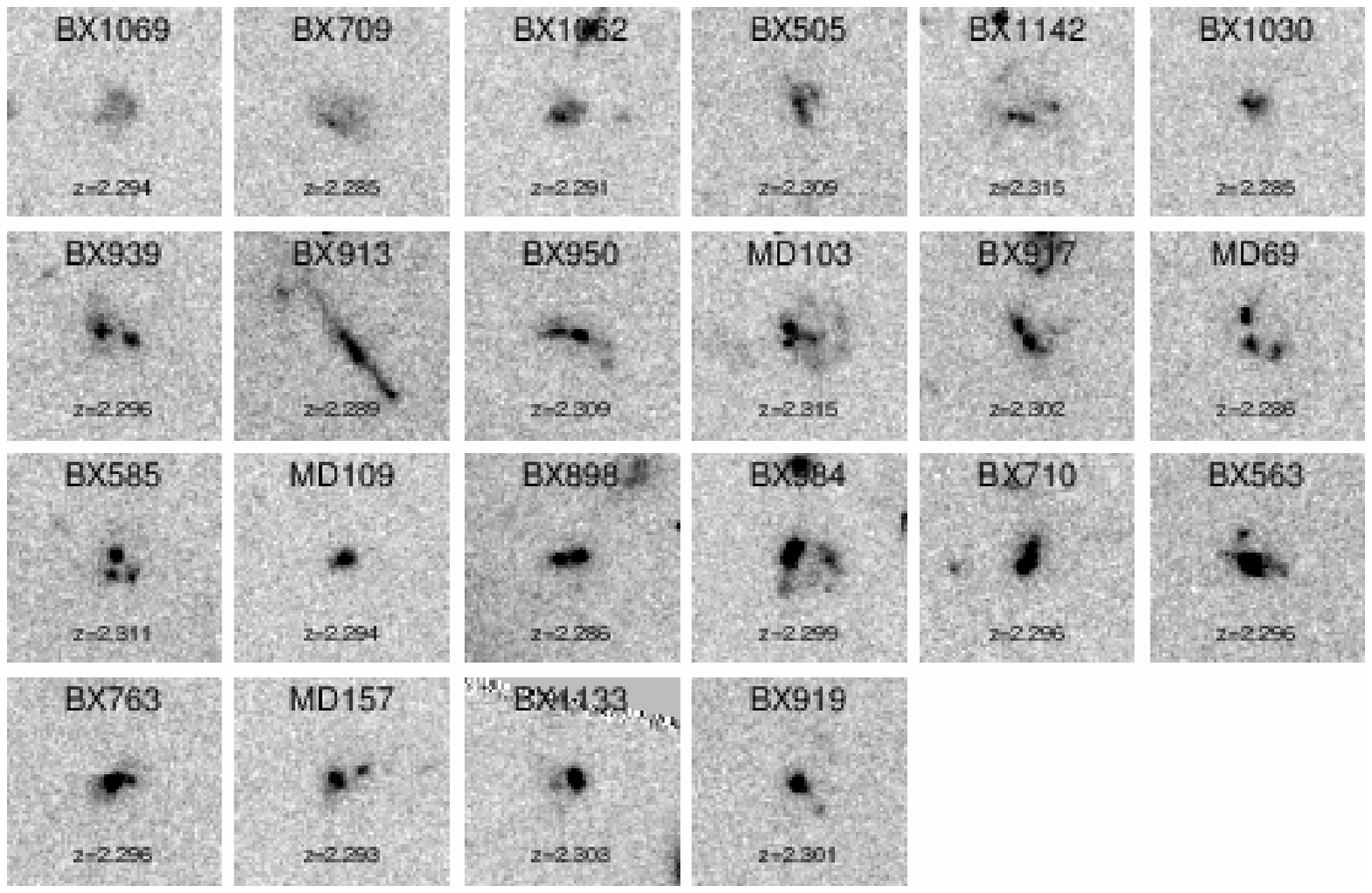}
\caption{\emph{HST}/ACS images of galaxies spectroscopically confirmed to be in the protocluster, sorted in order of increasing gini.  Each thumbnail is $4^{\prime\prime} \times 4^{\prime\prime}$ in size.  Images are centered at the centroid of the $\mathcal{R}$-band isophotes.  The elongated morphology of BX 913 may be due to do lensing by the foreground X-ray cluster RXJ1701.3+6414 at $z = 0.453$ \citep{mullis2003}.} 
\label{fig:spike}
\end{figure*}

\subsection{\emph{HST}/ACS Observations}
We imaged the Q1700 field with the \emph{HST}/ACS using the F814W filter.  For a protocluster mean 
redshift of $z=2.300$, the pivot wavelength of the filter corresponds to a rest-frame wavelength of 
$\lambda \sim 2500 \mbox{\AA}$, and the FWHM of the point-spread function of $0\secpoint12$ 
corresponds to $\approx 1$ kpc at $z=2.3$.  In order to include as many of the spectroscopically 
confirmed galaxies as possible, we covered the region nearest the HS1700+643 quasar with  four 
pointings (Figure \ref{fig:foot}).  Each pointing was imaged over the course of five orbits of 
nearly equal exposure time, for a total exposure time of 12520 seconds.  This corresponds to a 
sensitivity of 29.0 AB magnitude for a 1$\sigma$ surface brightness fluctuation in a 1 
$\Box^{\prime\prime}$ aperture, and a 28.4 AB magnitude depth for a 10$\sigma$ point source in a 
$0\secpoint1$ radius circular aperture.  We used the MultiDrizzle script \citep{koekemoer2002} to 
clean, sky-subtract, and drizzle the flat-fielded data products from the ACS CALACS software 
pipeline.  In order to align the ACS image to the ground-based coordinate system, we used $\approx 
100$ bright objects on each pointing to create a map between the ACS and ground-based coordinate 
systems. We then corrected the ACS image world coordinate system accordingly.

\subsection{Sample Selection}
 High redshift photometric candidate galaxies were selected using the BX/MD/C/D/M photometric color 
selection schemes of \citet{steidel2003} and \citet{adelberger2004a}.  A total of 1472 ``BX'', 238 
``MD'', 81 ``C'', 74 ``D'', and 45 ``M'' candidate galaxies were identified in the optical 
ground-based observations of the Q1700 field, for a total of 1910 optically selected galaxy 
candidates \citep{steidel2004,steidel2005}. Of primary interest to us for our environmental study 
of the $z=2.300$ galaxy redshift spike (``protocluster'') are the BX/MD objects, for which the mean 
redshifts are $z=2.20$ and $z=2.79$ respectively \citep{steidel2003,adelberger2004a}, in contrast 
to the mean redshift $z \approx 3$ for the combined C/D/M sample.  As discussed in Section 
\ref{sec:obs}, 141 of these galaxies have secure redshifts, of which 25 are in the overdensity.  In 
the near-infrared field, we identified 75 DRGs, 11 of which are also classified as BX or MD 
objects.  Of the DRGs that are also classified as BX or MD objects, 7 have spectroscopic redshifts, 
of which two are in the overdensity.  Figure~\ref{fig:foot} illustrates the footprints of all the 
observations and all of the galaxy samples.
\newline\indent
A total of 22 of 25 BX/MD objects whose spectroscopic redshifts place them in the protocluster fell 
on the footprint of our ACS observations.  Additionally, 72 optically selected objects with 
redshifts $z > 1.7$ but not within the protocluster were observed by ACS, as were 234 BX/MD/C/D/M 
objects without spectroscopic redshifts (of which 223 are not obviously stars or large foreground 
galaxies) and 43 DRGs.  Figures \ref{fig:spike}--\ref{fig:drg} show images of the spectroscopically 
confirmed optically selected galaxies, and all the DRGs.  All seven of the DRGs that also satisfy 
optical selection criteria and have spectroscopic redshifts are on the ACS pointings, as are three 
of the four DRGs that are also BX or MD objects that do not have secure redshifts.  A summary of 
the ACS galaxy samples can be found in Table 1.

\begin{figure*}
\epsscale{1.15}
\plotone{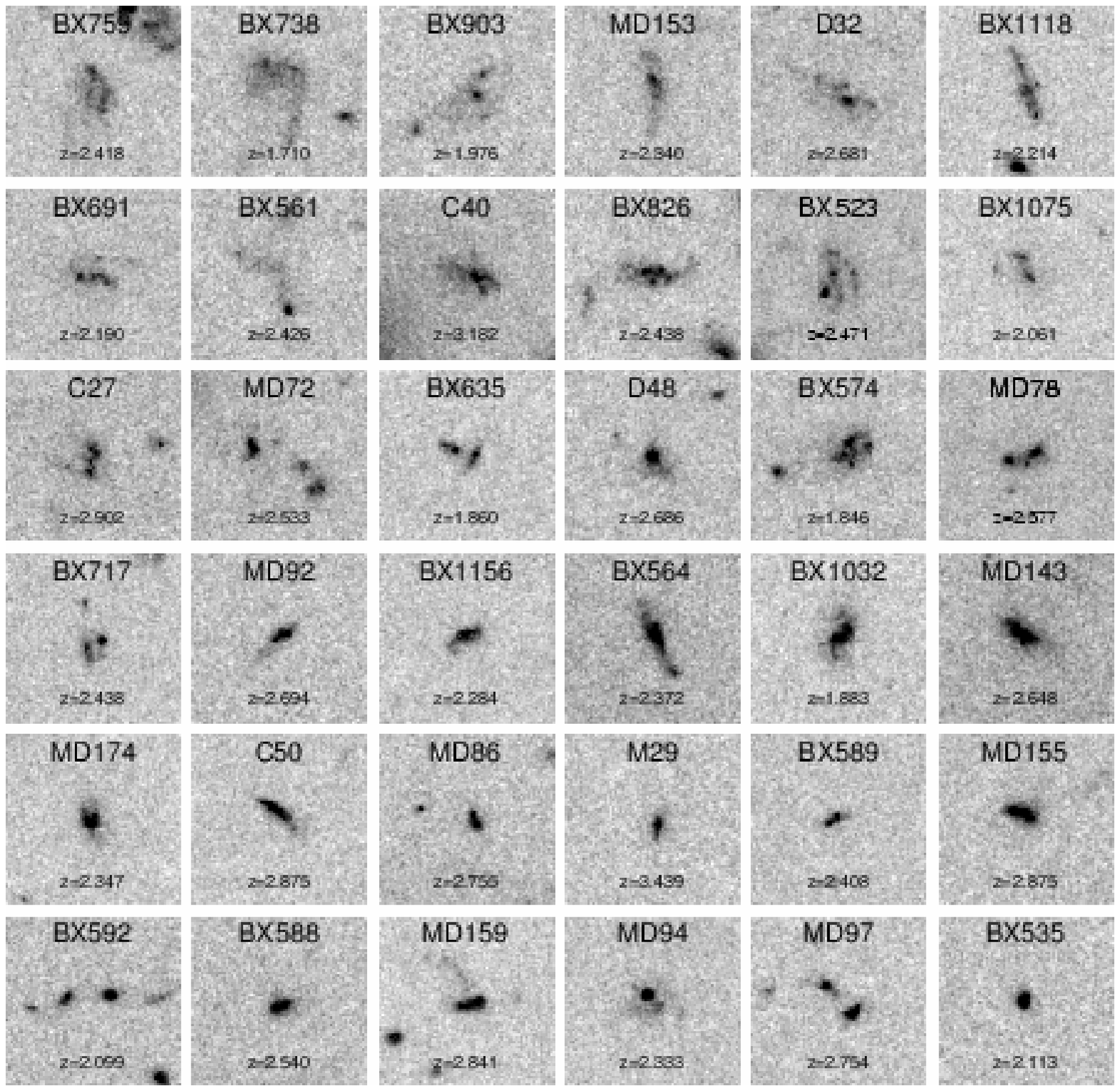}
\caption{\emph{HST}/ACS images of galaxies spectroscopically confirmed not to be in the protocluster.  As in Figure \ref{fig:spike}, galaxies are sorted in order of increasing gini, spanning the range of 0.245 to 0.449. Images are centered on the centroids of the $\mathcal{R}$-band isophotes.  See Figure \ref{fig:no2} for ginis in range 0.449 to 0.790.  }
\label{fig:no1}
\end{figure*}

%\clearpage
\begin{figure*}
\epsscale{1.15}
\plotone{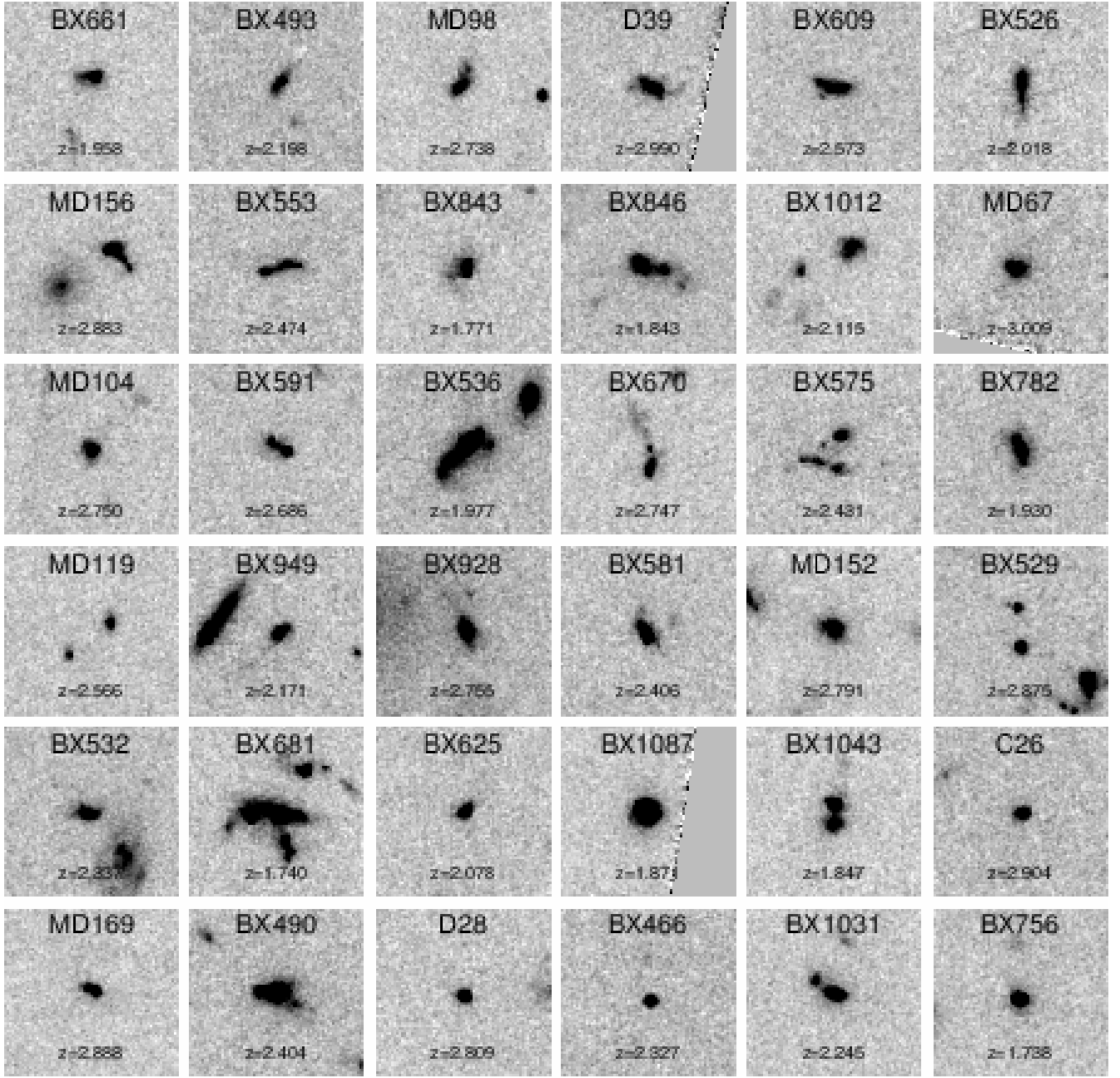}
\caption{Continuation of Figure 3.  \emph{HST}/ACS images of galaxies spectroscopically confirmed to NOT be in the redshift spike.  Galaxies are sorted in order of increasing gini, spanning the range $0.449 \leq G \leq 0.790$.  See Figure \ref{fig:no1} for galaxies with ginis in the range 0.245 to 0.449.}
\label{fig:no2}
\end{figure*}

%\clearpage
%%% DRG thumbnails
\begin{figure*}
\epsscale{1.15}
\plotone{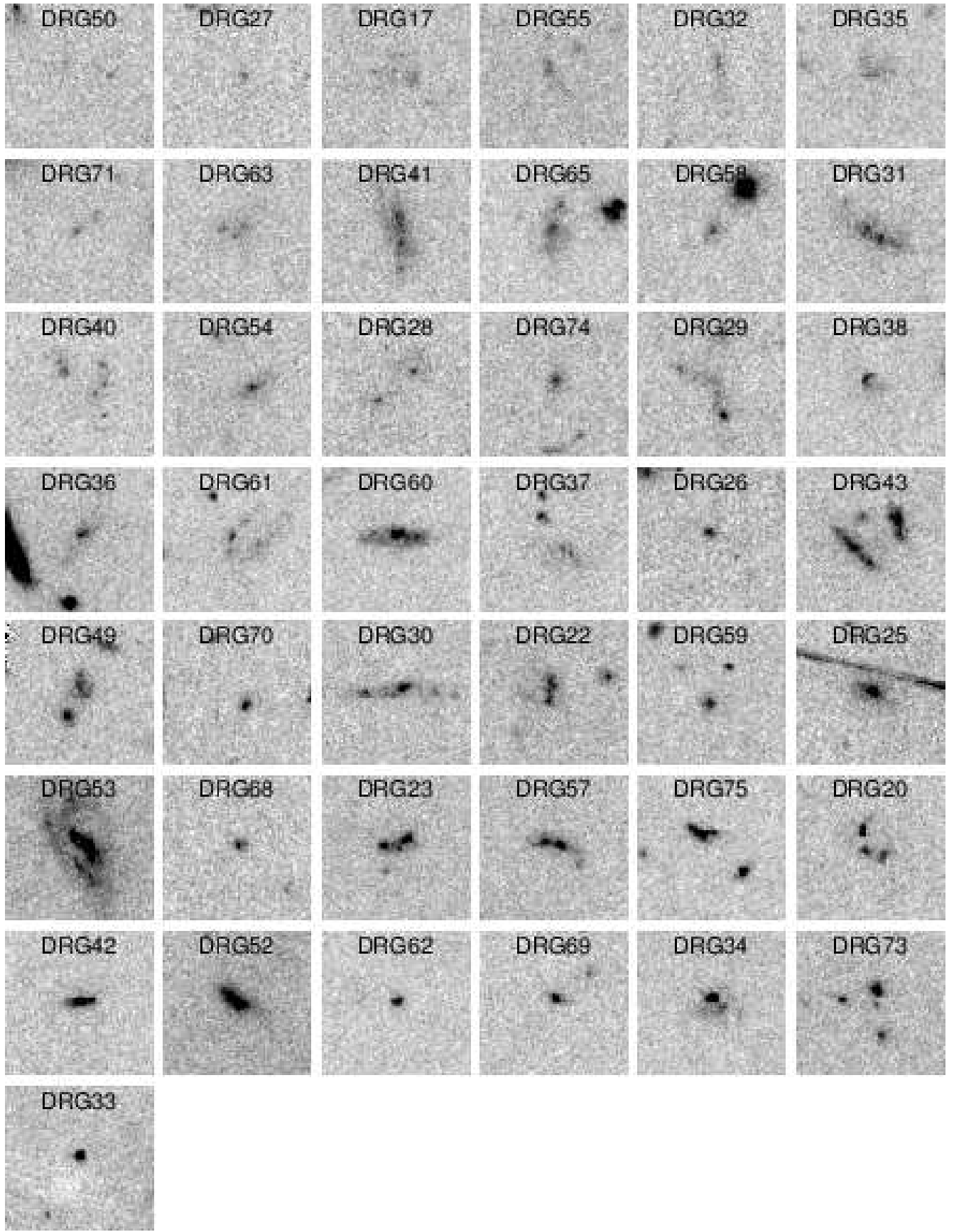}
\caption{\emph{HST}/ACS images of objects satisfying the $J-K_s > 2.3$ DRG color criterion.  Objects are sorted in order of increasing gini.  Images are centered at the centroid of the $K_s$-band isophotes.}
\label{fig:drg}
\end{figure*}

\begin{deluxetable}{lc}
%\tablenum{tab:samp}
\tablewidth{0pc}
\footnotesize
\tablecaption{Galaxy Samples}%\label{tab:samp}}

\tablehead{
\colhead{Sample} &\colhead{Number[SEDs]\tablenotemark{a}}
}

\startdata
BX/MD spike &22[15] \\
BX/MD non-spike &63[45] \\
\hline
total BX/MD w/redshift &85[60] \\
	& \\
BX/MD no redshift &197 \\
\hline
total BX/MD	&282 \\
	& \\
DRG &43\tablenotemark{b} \\
%Lyman-$\alpha$ emitter &17 \\

	& \\
LBG (C/D/M) with redshift &9[2] \\
LBG (C/D/M) no redshift &26 \\
\hline
total LBG (C/D/M) &35 \\

\enddata

\tablenotetext{a}{SED fits were obtained for objects with near- and mid-IR imaging and spectroscopic redshift information.  These are the objects used in the comparison of physical parameters and morphology.}
\tablenotetext{b}{Some DRGs are also selected optically (10 total) and have spectroscopic redshifts (7 of 10); however, redshifts are not available for most of the sample.}

\end{deluxetable}

\section{Morphological Parameters}
\subsection{Parameter Choice}
Galaxies at $z \sim 2$ are not classified easily by the Hubble sequence.  As demonstrated in 
Figures 2$-$5 and as noted in previous studies of the rest-frame UV and optical morphologies of 
$z>2$ galaxies \citep[e.g.,][]{dickinson2000, elmegreen2004, lotz2006, law2007}, high redshift 
galaxies tend to be clumpy and highly irregular, with no strong evidence for spiral or elliptical 
types.  Therefore, there have been several attempts to define a set of quantitative measures of 
morphology.  Some \citep[e.g.,][]{ravindranath2006,zirm2006} have used circularly symmetric 
Sersi{\' c} fits to quantify the surface brightness distribution.  Others have used the \emph{CAS} 
(concentration \emph{C}, asymmetry \emph{A}, clumpiness \emph{S}) system 
\citep{kent1985,schade1995,bershady2000,conselice2003}.  However, both the Sersi{\' c} fits and the 
\emph{C} and \emph{A} parameters of \emph{CAS} are defined assuming that elliptical isophotes 
provide a good description of the objects in question.  These parameters are also extremely 
sensitive to the precise location of the ``center'' of the galaxy.  Given the highly irregular 
nature of high redshift galaxy morphologies, it is unclear if elliptical, centroid-sensitive fits 
describe the galaxies in a robust and meaningful way.  
\newline\indent
Instead, we adopt a set of non-parametric coefficients to estimate galaxy morphologies.  These 
coefficients have the advantage of not enforcing symmetries on the galaxy light distribution.  The 
parameters we use are: size, gini, and multiplicity.
\subsubsection{Size}
%To find the size:
%Da in cosmolib.c outputs in Mpc.
%Plug into David's formula in his paper

The simplest way to quantify the size of a galaxy at high redshift is by its projected (angular or 
physical) surface area.  In particular, since the angular diameter distance in our cosmology is $< 
10$\% smaller at the upper end of the BX/MD spectroscopic sample redshift distribution than at the 
lower end, we use the angular projected surface area as the measure of galaxy size.  This is also a 
pragmatic choice given that we do not have spectroscopic redshifts for some of our galaxy samples.  
For simplicity, we define size
\begin{eqnarray}
	\mathit{I} = N_{pix},
\end{eqnarray} 
where $N_{pix}$ is the number of pixels associated with the galaxy.  To convert to angular units, 
one must multiply \emph{I} by the angular size of an ACS pixel ($(0\secpoint05)^2$).  A typical 
size of 250 ACS pixels corresponds to 0.625 $\Box^{\prime\prime}$, or $43$ kpc$^2$ at $z=2.3$.  

\subsubsection{Gini}
An important measure of galaxy morphology is the distribution of light emission, in the sense that 
one would like to know if the galaxy surface brightness is roughly constant, or if there are large 
variations in surface brightness.  For galaxies with a high degree of symmetry (for example, 
galaxies that are easily classified by the Hubble sequence), measures such as the Sersi{\' c} index 
or the concentration parameter \emph{C} (proportional to the logarithm of the ratio of the radius 
containing 80\% of the light in the galaxy to the radius containing the inner 20\%) indicate the 
strength of variations in surface brightness.  However, since high redshift galaxies are irregular, 
we instead adopt the gini parameter
\begin{eqnarray}\label{gini}
	\mathit{G} = \frac{1}{\bar{X}N_{pix} (N_{pix} - 1)} \sum^{N_{pix}}_{i=1} (2i - N_{pix} -1)X_i,
\end{eqnarray}
where $\bar{X}$ is the mean sky-subtracted flux per pixel, $X_i$ is the sky-subtracted flux in 
pixel $i$, and the pixels have been sorted in order of increasing flux.  This parameter, originally 
used in economics to describe the distribution of wealth in a population, was first used in an 
astronomical context by \citet{abraham2003}, and its usefulness in the context of galaxy morphology 
has been examined by \citet{lotz2004,lotz2006} and \citet{law2007}.  
\newline\indent
Gini is defined to have a value $G = 0$ if the flux is uniformly distributed among pixels, and 
$G=1$ if one one pixel hoards all the flux in the galaxy.  Thumbnails of the spectroscopic BX/MD 
and the DRG samples are sorted in order of increasing gini (Figures 
\ref{fig:spike}--\ref{fig:drg}).  Note that objects with the same gini may span the range of 
diffuse objects with multiple small knots to those with a single bright central region.  Therefore, 
unlike the Sersi{\' c} index or \emph{C}, gini can accommodate a wide range of spatial light 
distributions.  

\subsubsection{Multiplicity}
\begin{figure}
\epsscale{1.15}
\plotone{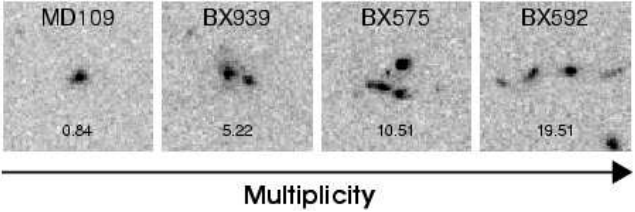}
\caption{Sample galaxies (MD 109, BX 939, BX 575, and BX 592) demonstrate the range and interpretation of multiplicity.}
\label{fig:mult}
\end{figure}

In addition to describing how light is distributed among pixels, one would like to characterize the 
shape or clumpiness of the galaxy light distribution.  We use the multiplicity ($\Psi$) parameter 
introduced by \citet{law2007} as a measure of this property.  Multiplicity is analogous to the 
total potential energy of a gravitational system.  The total ``potential energy'' of the flux 
distribution can be described as
\begin{eqnarray}
	\psi_{tot} = \sum^{N_{pix}}_{i = 1} \sum^{N_{pix}}_{j=1,j\neq i} \frac{X_i X_j}{r_{ij}},
\end{eqnarray}
where $r_{ij}$ is the distance (in pixels) between the $i$th and $j$th pixels.  If one were to 
reorder the pixels so that they were in the most compact configuration (i.e., with the brightest 
pixel at the center, and the next brightest pixels forming a ring around that pixel, and so forth), 
this configuration would have the greatest ``total potential energy'':
\begin{eqnarray}
	\psi_{compact} = \sum^{N_{pix}}_{i = 1} \sum^{N_{pix}}_{j =1,j\neq i} \frac{X_i X_j}{r^{\prime}_{ij}},
\end{eqnarray}
where $r^{\prime}_{ij}$ is the distance between the $i$th and $j$th particles in the new 
configuration.  The multiplicity is then defined to be
\begin{eqnarray}
	\Psi = 100 \log_{10} \left[ \frac{\psi_{compact}}{\psi_{tot}} \right].
\end{eqnarray}
\newline\indent
The range of $\Psi$ depends somewhat on the method used to assign pixels to galaxies (see Section 
\ref{sec:pix}).  Generally, objects with $\Psi \lesssim 2$ are bulge-like and objects with $\Psi < 
5$ tend to be dominated by one main clump. Objects with high multiplicity ($\Psi \gtrsim 15$) tend 
to have many almost equally bright sources that are widely separated.  Galaxies with intermediate 
multiplicity can be highly elongated, or have several clumps that are close together.  In general, 
galaxies that consist of several clumps of uneven luminosity will have lower multiplicity than a 
galaxy for which all the clumps are of approximately equal luminosity.  Figure \ref{fig:mult} 
demonstrates galaxies that are typical of their multiplicities.

%%% Multiplicity gradient

\subsection{Pixel Selection}\label{sec:pix}
A major problem in characterizing the morphologies of high redshift galaxies is determining which 
pixels are associated with the galaxy, and how the pixel selection affects the quantitative 
measures of morphology.  \citet{lotz2004} select pixels by first determining the ellipticity and 
Petrosian radius of a galaxy, using elliptical apertures.  They then convolve their data within the 
Petrosian ellipse with a gaussian, and select all pixels within the Petrosian ellipse that meet a 
surface brightness cutoff.  As discussed in \citet{law2007}, the caveats associated with this 
approach include the fact that it assumes symmetry for a population of highly irregular objects, 
and it tends to associate many sky pixels with the galaxy.
\newline\indent
\citet{law2007} use a method that does not depend on the radial symmetry of the object.  An initial 
estimate of the galaxy centroid is obtained from ground-based imaging.  The galaxy centroid is then 
recalculated, based on the light distribution in the ACS images.  Next, an aperture of 
$1\secpoint5$ radius is used to search for pixels that are associated with the galaxy.  Pixels are 
assigned with a galaxy if they meet a redshift-dependent surface brightness threshold.  The 
redshift-dependent surface brightness threshold scales with cosmological dimming, and its purpose 
is to prevent bias in morphological measurements.  This threshold is defined as $n(z)\sigma$, where 
$\sigma$ is the standard deviation in the sky background.  \citet{law2007} set the threshold for 
the highest redshift galaxy to $n(z_{high}) = 3$, where $z_{high}= 3.4$.  Then, since surface 
brightness dims as $(1+z)^{-3}$ for a fixed passband, they set the surface brightness threshold 
\begin{eqnarray}
	n(z) = 3 \left[ \frac{1+z_{high}}{1+z} \right]^3 .
\end{eqnarray}
The advantage of this surface brightness threshold method is that the geometry of the object does 
not matter.  This disadvantage of this method, for our purposes, is that that it requires redshift 
information, which is missing for many of our galaxies, in particular, the majority of the DRGs.  
\newline\indent
For the current analysis, we adopt a pixel selection method that is redshift independent and 
includes all the pixels associated to the galaxy without including sky pixels.  We use a redshift 
independent selection method for two reasons.  First, given the poor constraints on photometric 
redshifts, it does not make sense to use a redshift-dependent threshold based on photometric 
redshifts for our samples without spectroscopic information.  Secondly, the redshifts of our 
spectroscopic sample span a smaller range than those of \citet{law2007} and \citet{lotz2006}. 
\citeauthor{lotz2006} examine two samples with mean redshifts $z\sim 1.5$ and $z\sim 4$, while 
\citeauthor{law2007} break up their $1.8\leq z \leq 3.4$ sample into two samples with mean 
redshifts of $z\sim 2$ and $z\sim 3$. More than 75\% of the galaxies in our BX/MD sample fall into 
the lower-redshift bin from \citet{law2007}. Furthermore, though they span a larger redshift range 
than confirmed protocluster members, galaxies in the Q1700 field that are not contained within the 
redshift spike have a similar mean redshift, $\bar{z} = 2.34$.  The similarity in the mean 
redshifts is due to the fact that $z=2.3$ lies near the peak of the BX/MD redshift selection 
function \citep{steidel2004}.  Therefore, it is less crucial in the current study to account for
cosmological dimming in the surface brightness threshold.
\newline\indent
The first step in selecting pixels associated with a particular galaxy is to define the region in  
which flux could reasonably be associated with the galaxy.  Like \citet{law2007}, we begin by
recalculating the galaxy centroid in
order to center the pixel selection region.  The centroid is calculated based on the first-order 
moment of
light on the ACS images, usually calculated within a 30-pixel ($1\secpoint5$) radius circle centered
on the $\mathcal{R}$
centroid.  We use a different aperture within which to calculate the centroid if the multiband 
ground-based images indicate that there is
another object (either a large foreground galaxy or galaxy with different colors than the object in
question) within the
default $1\secpoint5$
aperture, or if the $\mathcal{R}$-band isophote has a linear extent $>3^{\prime\prime}$.  In those
cases, the size of the
centroid-finding aperture is tailored to avoid contaminants and completely surround the
$\mathcal{R}$-band detection.  In
some cases, a single ground-based detection will encompass several clumps of light with separations
of $\sim
1^{\prime\prime}$ (i.e., roughly the size of the
ground-based seeing), which corresponds to a projected physical separation of 8 kpc at $z=2.3$.  It
is highly unlikely
that such small separations
are the result of a projection effect, so we consider all clumps that fall within the ground-based
detection isophote to be
associated with
each other.  Therefore, we calculate the centroid for all patches of light that lie in the
ground-based detection.
\newline\indent
Next, we define an aperture about the newly calculated centroid within which we search for pixels  
associated with the galaxy.  As in \citet{law2007}, the default aperture size is a $1\secpoint5$
radius circle.  However, as in our search for the galaxy centroid, we must customize the aperture
size in some cases.  In those cases, we use the ground-based $\mathcal{R}$-band images as a guide,
since the seeing-limited $\mathcal{R}$-band detections provide an approximate upper limit to the
extent of the galaxy.  If the $\mathcal{R}$-band detection extends beyond $1\secpoint5$ ($\sim 12$
kpc in projection) from the center of the object, we use a circular aperture on the ACS image that
encompasses the $\mathcal{R}$-band detection.  If there are foreground galaxies or objects that have
significantly different colors than the object in question that lie within
the default aperture, which occurs for $\sim 15 \%$ of the spectroscopic sample, we use circular
apertures that are small enough to avoid contaminants. It was only necessary to use a non-circular
aperture twice (for the photometric BX/MD sample) in order to avoid contaminants.  In these cases,
the elliptical apertures were aligned to the long-axis of the galaxy and made large enough to
encompass the whole galaxy, but small enough to avoid contaminants.  The size of those
elliptical apertures was also adjusted so that the
total area within the aperture was not signifcantly different from the area of the default aperture.

%\clearpage
%%%Selection contours
\begin{figure}
\epsscale{1.15}
\plottwo{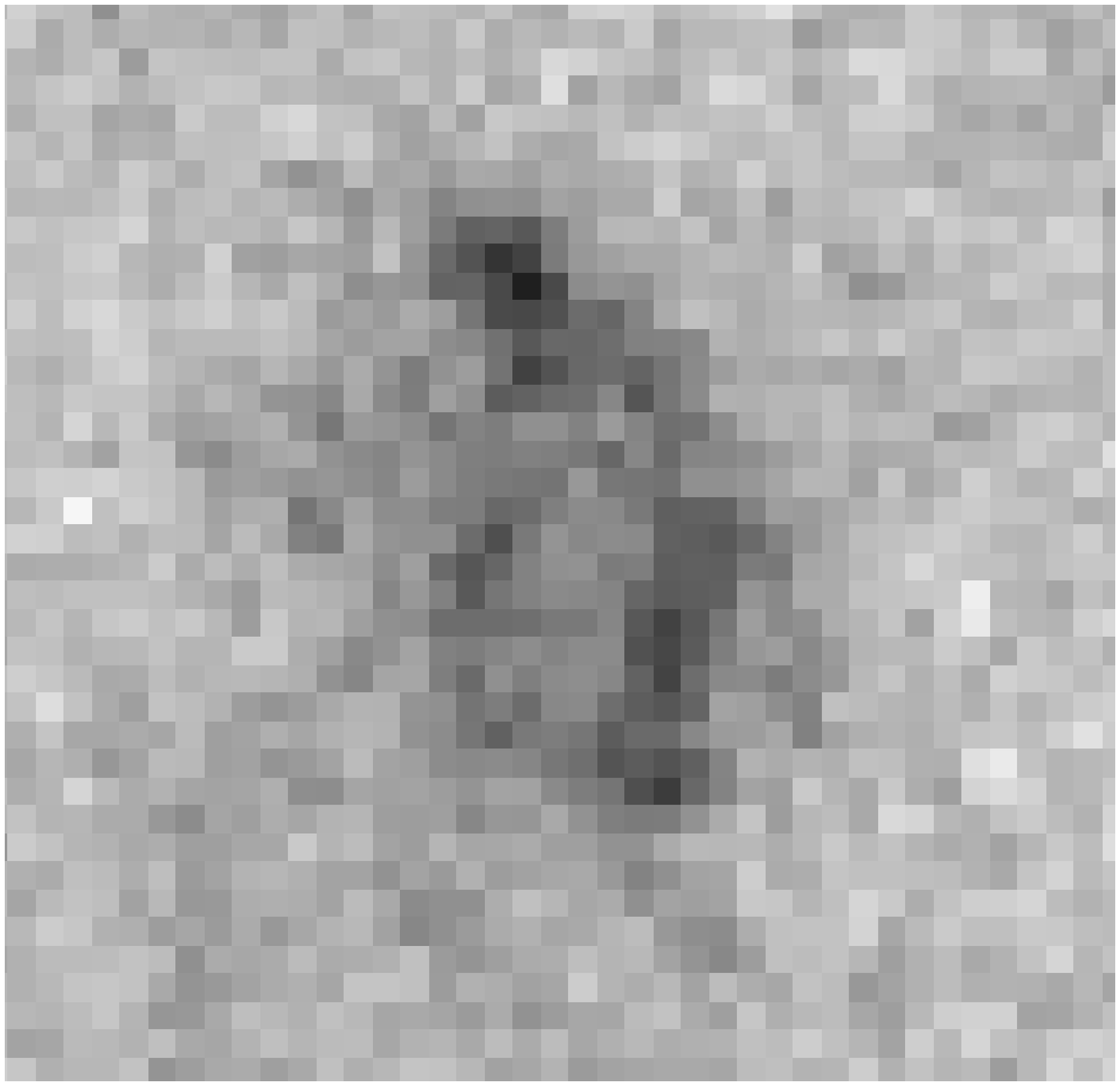}{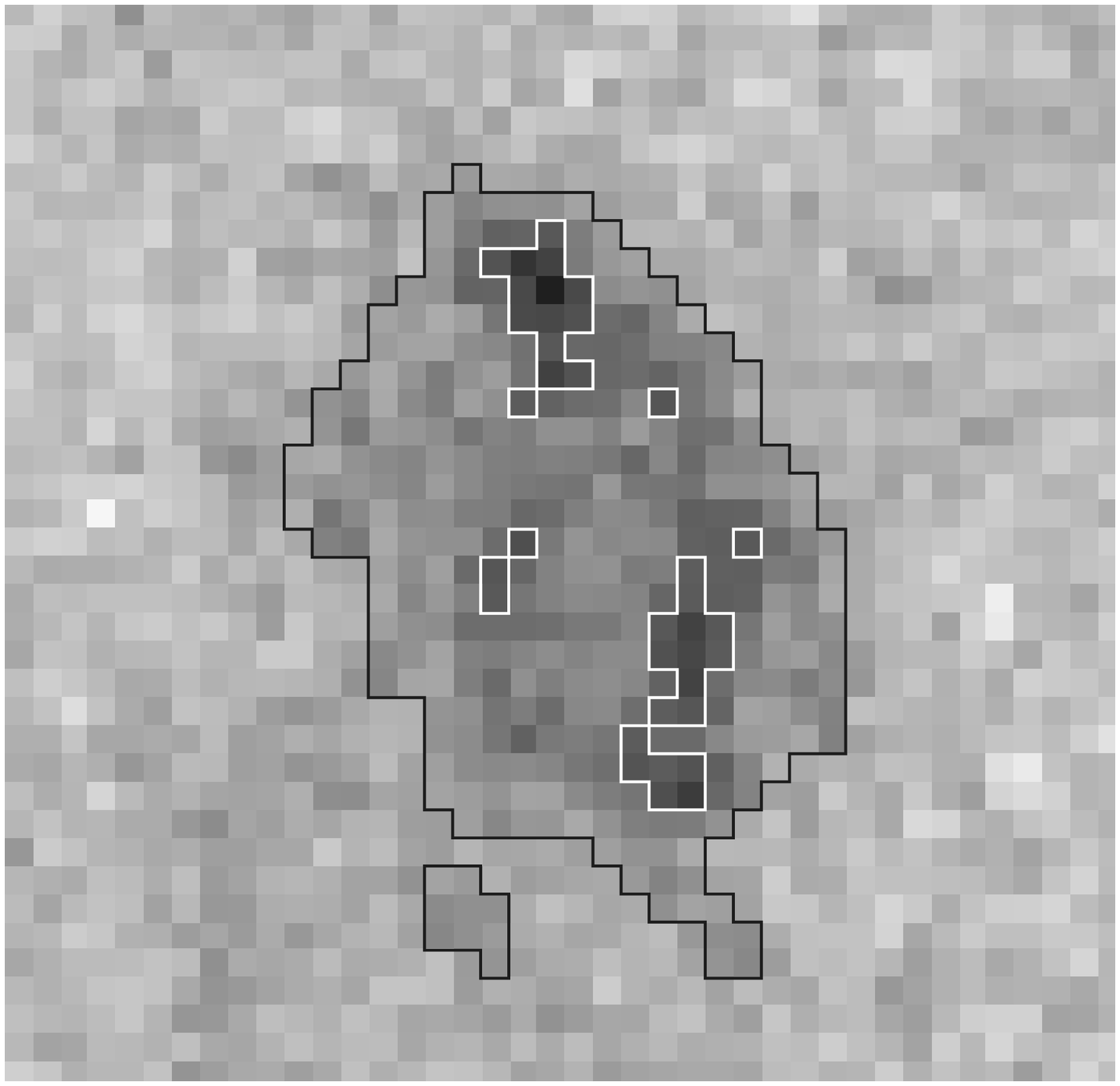}
\caption{Pixel selection for BX 759 ($z=2.418$), which would be classified as ``nebulous'' in Law et al. (2006). \emph{Left:} BX 759 without pixel selections marked.  Image size is $2^{\prime\prime}\times 2^{\prime\prime}$.  \emph{Right:} White contours mark pixel selection according to the redshift-dependent pixel selection scheme of Law et al. (2007; Scheme B).  This corresponds to selecting all pixels $ > 8\sigma$ above the sky at $z=2.418$ such that galaxies at $z_{high}=3.4$ are selected with a $3\sigma$ surface brightness threshold.  Black lines delineate the regions which are associated with the galaxy using the fiducial pixel selection scheme.}
\label{fig:contour}
\end{figure}

%\clearpage
%%%comparison
%%%
%%%col 1: size
%%%col 2: gini
%%%col 3: multiplicity
%%%col 4: flux
%%%row 1: David's z-dependent 30-pix
%%%row 2: David's aperture
%%% all plotted vs. G1 T2
\begin{figure*}
\epsscale{1.15}
\plotone{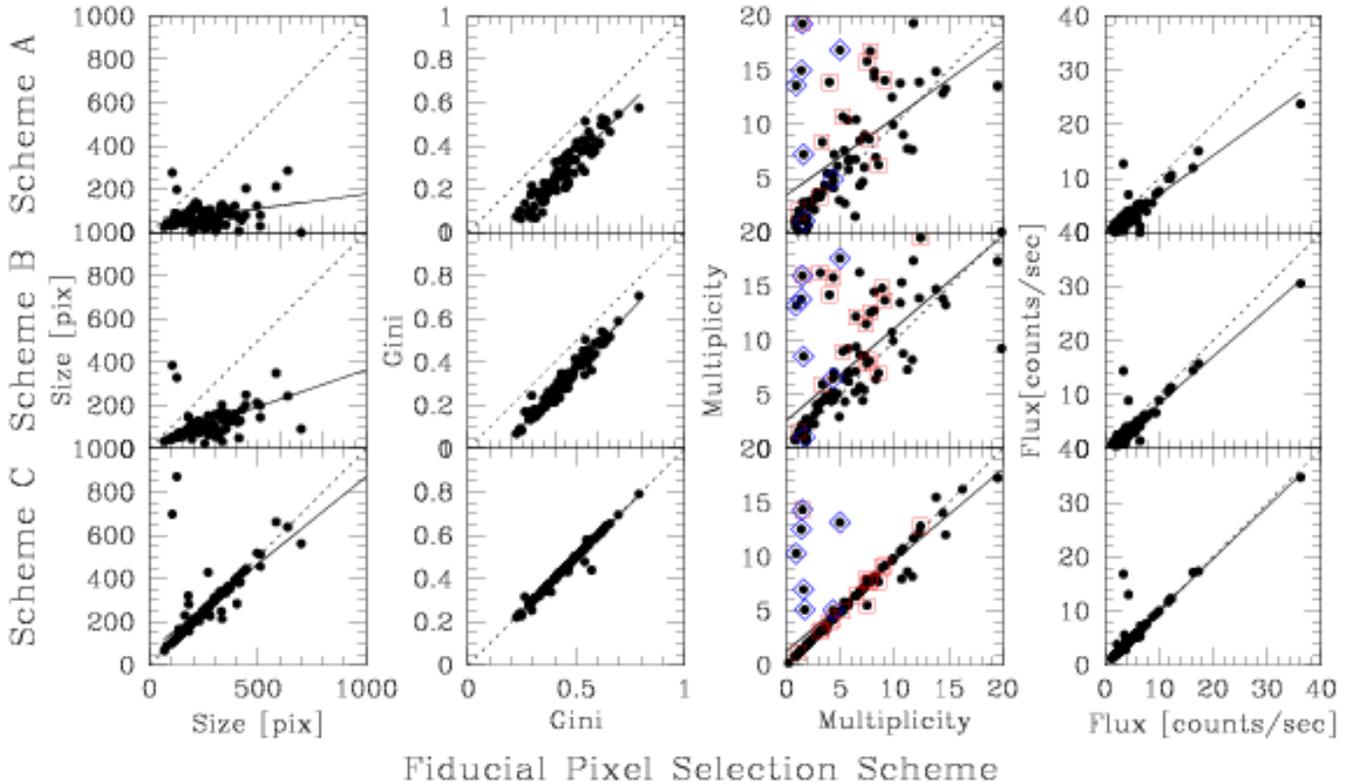}
\caption{Morphological parameters as a function of pixel selection.  Blue diamonds mark galaxies for which the 30 pixel aperture captures flux from nearby objects.  Red squares mark galaxies that classified as ``nebulous" by \citet[][$G<0.15$ in their scheme]{law2007} for which multiplicity is ill-defined.}
\label{fig:compare}
\end{figure*}

%\newline\indent
As in \citet{law2007}, we perform local sky-subtraction for each object to account for any 
imperfections in the MultiDrizzle sky subtraction.  We calculate the sky background in an annulus 
about the detection aperture, typically 6 $\Box^{\prime\prime}$ (or $\sim$2400 pixels) in total 
surface area.  We use a sigma-clipping algorithm to exclude any pixels that vary by more than 
3$\sigma$ about the sky level from the calculation of the sky background.
\newline\indent
In order to determine which pixels in the detection aperture are associated with the galaxy, we 
convolve the data with a circularly symmetric gaussian with a one-pixel standard deviation, 
truncating the gaussian four standard deviations from the center, using the IRAF tool ``gauss.''  
This allows us to exclude sky pixels from our segmentation map, since smoothing reduces the 
significance of isolated bright pixels.   We select pixels to be associated with the galaxy if the 
smoothed flux is more than 2$\sigma$ above the sky, with $\sigma$ calculated from the raw data.  We 
then use the raw sky-subtracted flux in those pixels for our morphological analysis. In the rest of 
the paper, we refer to this pixel selection scheme as the ``fiducial'' scheme.  Figure 
\ref{fig:contour} shows pixels selected with both our fiducial method and the method of \citet{law2007}.

%\newline\indent
In the interest of determining the robustness of morphological parameters derived using our pixel 
selection scheme, we compare the values of size, gini, multiplicity, and the total flux using our 
pixel selection with results from various other pixel selection methods (Figure \ref{fig:compare}).  
Points on Figure \ref{fig:compare} correspond to all high redshift galaxies with spectroscopic 
redshifts, including C/D/M galaxies, $z\sim 3$.  We consider three reasonable schemes for selecting 
pixels.  In Scheme A, all pixels with fluxes satisfying a $5\sigma$ threshold on the unsmoothed 
imaged within a $1\secpoint5$ radius of the centroid are associated with the galaxy.  This allows 
us to evaluate the effects of smoothing for constant surface brightness selection schemes.  In 
Scheme B, we use the redshift-dependent pixel selection scheme of \citet{law2007} to associate 
pixels with galaxies.  This allows us to examine the effects of a redshift-dependent surface 
brightness threshold.  In Scheme C, we use the fiducial scheme with the one modification that we 
fix the detection aperture to $1\secpoint5$, so that we can isolate the effects of aperture size on 
the morphological parameters.  In Figure \ref{fig:compare}, the dotted line in each panel is the 
line of equality ($y=x$), and the solid line is the least-squares fit to the data.  
\newline\indent
Since we are interested in the differences of the morphologies of different samples 
(spectroscopically confirmed protocluster and ``field'' BX/MDs, BX/MDs without redshifts, and 
DRGs), it is less important that the quantitative measures of morphology be absolutely calibrated 
as a function of pixel selection than it is that they be robust in a differential way.  In other 
words, we would like a strong correlation to exist between morphological parameters calculated with 
two different pixel selection schemes, but it is not necessary that the parameters take the same 
value.  We see that the sizes measured with different pixel selection schemes are well-correlated 
(see Figure \ref{fig:compare}), as are the ginis and fluxes.  This is especially true for Scheme C, 
since it differs from the fiducial scheme only in the size of the detection aperture.
\newline\indent
Multiplicity shows much more scatter than the other parameters.  However, most of the scatter comes 
from only two effects.  Points in Figure \ref{fig:compare} outlined with blue diamonds represent 
galaxies for which the $1\secpoint5$ detection aperture includes contamination from other galaxies 
that are nearby in projection.  Because multiplicity is sensitive to the spatial distribution of 
light, the effect of the contaminants is to systematically increase the multiplicity.  Points 
outlined with red squares indicate galaxies that have $G < 0.15$ as calculated in Scheme B 
\citep[i.e., in the fashion of ][]{law2007} but have high multiplicity.  \citet{law2007} classify 
galaxies with these parameters as ``nebulous,'' and comment that multiplicity is ill-defined for 
these galaxies.  Since they tend to have low surface brightness, they are quite sensitive to the 
pixel selection method.  We have designed our fiducial scheme to capture as much information from 
these objects as possible (see Figure \ref{fig:contour} for an example of which pixels the fiducial 
scheme and Scheme B select for a ``nebulous'' galaxy).
\newline\indent
Therefore, we can conclude that, for the most part, our set of morphological parameters is robust 
in a differential sense to pixel selection.  We use our fiducial pixel selection method throughout 
the paper, but will occasionally check results from the fiducial scheme against those from the 
other schemes. 
\newline\indent
Even for a fixed pixel selection technique, there are uncertainties
in morphological parameters due to the finite signal-to-noise (S/N)
of the object detections. We assess such uncertainties
by running simulations of our morphological analysis procedure
with the fiducial pixel selection technique.
Using pixel flux distributions drawn from our actual sample of objects, 
spanning a range of size, gini, and multiplicity, we add copies
of each object light distribution to 100 random locations in
the ACS image, and recover the morphological parameters 
for simulated objects in relatively clean locations
(i.e. objects whose fluxes are not significantly contaminated
by nearby bright objects). This procedure indicates that
the typical size uncertainties are $5-10$\%, the typical
uncertainties on gini values are $<5$\% (i.e. $\sigma(G)=0.01-0.02$), 
and the typical multiplicity uncertainties are $5-10$\%. For the faint
DRGs in our sample, the uncertainties are larger, at the level of
$\sim 15$\% for size, 10\% for gini (i.e. $\sigma(G)=0.02$),
and 30\% for multiplicity.

%3rd nearest neighbor surface density (Cooper MNRAS 370 (2006) 198), Kauffmann paper

\section{Spike vs. Non-Spike Objects: Morphology as a Function of Environment}
\begin{figure*}
\plotone{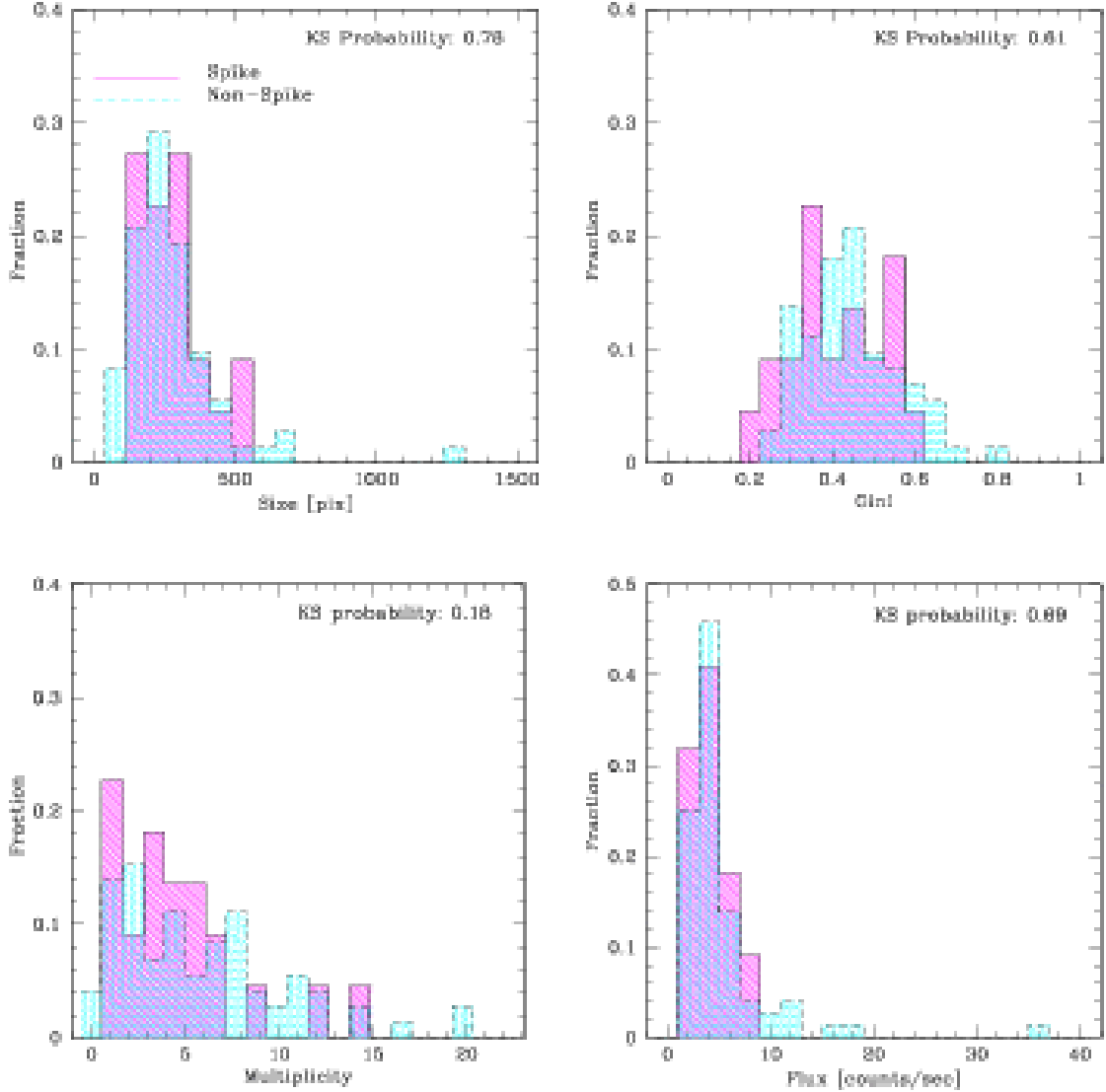}
\caption{Distributions of morphological parameters (size, gini, multiplicity, flux).  The morphology distributions of spike objects are denoted with magenta solid lines.  The distributions of the non-spike galaxies are denoted with cyan hatched lines.  The KS probability that the distributions of spike and non-spike objects are drawn from the same distribution is located in the upper right-hand corner of each plot.}
\label{fig:env}
\end{figure*}

Our primary interest in determining if there is a dependence of morphology on environment comes from the 
initial analysis of the Q1700 spike, which showed that certain properties of the spike galaxies were quite 
different from those not in the spike \citep{steidel2005}.  Empirically, galaxies within the spike have redder 
$\mathcal{R} - K_s$ colors than galaxies outside of the spike.  Stellar population synthesis modeling of the 
broadband colors of the spectroscopically confirmed BX/MD objects showed that, regardless of whether models 
were fit with a constant or exponentially declining star formation rate, galaxies contained in the spike are 
significantly older and more massive than those not in the spike.  These facts suggest that star formation 
history is already a function of environment at $z=2.3$.  
\newline\indent 
In this section, we will consider morphology as a function of environment; in the next section, we consider 
morphology as a function of physical properties (mass, age, etc.) and star formation history.  In this way, we 
can determine if environment, star formation history, and morphology are as tightly related at high redshift as 
they are in the current epoch.  As a cautionary note, we emphasize that our data probe rest-frame UV, not 
optical wavelengths, and so we are probing morphologies of the young star-forming regions of each galaxy 
instead of the stellar mass as a whole.  However, there is some evidence \citep{dickinson2000,zirm2006} that 
the rest-frame UV morphology is a reasonable tracer of optical morphology \citep[but see][]{toft2005}.
\newline\indent
For the analysis of the morphology distributions as a function of environment, we must first define what is 
meant by ``environment.''  In large surveys, such as SDSS and the DEEP2 galaxy redshift survey 
\citep{davis2003}, ``environment'' refers to the number density of galaxies, as calculated on Mpc scales.  In 
our context, environment refers not to the local number density of galaxies, but to the binary designation of 
whether or not a galaxy belongs to the protocluster.  We define environment in this way due to our small sample 
size.  Therefore, we separate our spectroscopic BX/MD sample into two subsamples.  The ``spike'' sample 
consists of 22 BX/MD galaxies with spectroscopic redshifts of $z = 2.300 \pm 0.015$.  As shown in 
Figure~1 of \citet{steidel2005}, the number
of galaxies in this small redshift interval
exceeds the average expected number according to the
BX/MD redshift selection function, by a factor of $\delta_g^z\sim 7$.
Typical redshift uncertainties of $\sigma_z\sim 0.002$
\citep{adelberger2005c} do not significantly
bias the estimate of the spike overdensity.  The 
``non-spike'' sample consists of 63 spectroscopically confirmed BX/MD galaxies with $\bar{z} = 2.34$.  Since we 
are interested in comparing the morphologies of galaxies at the same epoch but in different environments, we do 
not include the additional C/D/M galaxies because their high mean redshift $\bar{z}=2.94$.  The spectroscopic 
sample of non-spike BX/MDs serves as a good control sample due to the fact that its mean redshift is almost the 
same as that of the spike.  We calculate the size, gini, multiplicity, and total flux of each galaxy using the 
fiducial pixel selection method of Section 3.  
\newline\indent
To evaluate the differences between the spike and non-spike samples in the distributions of morphological parameters 
(Figure \ref{fig:env}), 
we perform a one-dimensional Kolmogorov-Smirnov (KS) test for each parameter.  As demonstrated in Figure \ref{fig:env} and 
Table 2, there is 
no statistically significant difference between the spike and non-spike distributions for each parameter.  In order to 
test the robustness of 
these findings, we perform KS tests for a variety of pixel selection methods.  Also listed in Table 2 are the KS 
probabilities that the spike 
and non-spike morphologies are drawn from the same distribution as calculated for Schemes A, B, and C.  For no pixel 
selection method are the 
spike morphological parameter distributions significantly different from the non-spike distributions.
\newline\indent
We also perform an analysis to determine if there are different proportions of ``bulge-dominated''
galaxies in the two samples, since,  in the local
universe, bulge-dominated systems (ellipticals and S0's) are preferentially found in dense
environments.  We define a galaxy to be bulge-like
if has a multiplicity $\Psi < 2$ and $G>0.5$, which means that it is a single compact system
with a sharply falling flux
distribution function.  While objects with $\Psi < 2$ are all spatially compact, they can have
flux distribution functions
characteristic of spheroids, exponential disks, or something in between.  The cut in gini
separates galaxies with steep
flux distribution functions (which would be characteristic of bulge-dominated systems) with those that have flux more
uniformly distributed across
pixels (typical of disk-like galaxies).   Separating the bulge population using cuts in two
of our morphological parameters is analogous to the classification of bulges in \citet{lotz2006}
using gini and the second moment of the light distribution.  This two-dimensional classification
of bulges is a cruder metric than, for example, the Sersi{\' c} index, and encodes no dynamical
information.  However, it is more useful if the bulk of the galaxy sample has irregular morphology, and does separate galaxies that appear to be compact and roughly axially
symmetric in projection from the more irregular galaxies.  Using this metric, we find that 3 of the 22
(14\%) spike galaxies are bulge-like, as are 11 of
the 63 (17\%) non-spike objects.  Given the small number of galaxies in each sample, the
difference in bulge fraction is not statistically significant.
\newline\indent
In order to explain the lack of differences in morphology between the spike and non-spike samples, there are four main points of interest.  
The first is that our sample size is fairly small.  One can detect large differences in distributions relatively easily with small samples, 
for example, the large differences in the mean stellar mass of the spike and non-spike galaxy samples.  However, it would be difficult to 
ascertain subtle differences with our small samples using the KS test, especially if the differences were mainly in the tail of the 
distributions, for which the KS test is not the optimal statistic.  The second point is that, with the F814W filter, we are probing a 
rest-frame wavelength that is in the UV.  If we are interested in morphology as a proxy for stellar mass distribution, it would be better to 
probe the galaxies at a rest-frame wavelength that is less sensitive to the brightest (youngest) stars and dust extinction.  While several 
groups \citep[e.g.,][]{dickinson2000,zirm2006} claim that differences in rest-frame UV and optical morphologies are small, these conclusions 
are based on very small samples imaged with the NIC3 camera on \emph{HST}, which has a coarser resolution ( $0\secpoint 26$) than that of ACS.  
Thirdly, we have analyzed the morphologies of galaxies that were photometrically selected to be star-forming and relatively unobscured.  
Hence, we are missing populations of old or obscured galaxies from our analysis; one might worry that we are doing the equivalent of measuring 
the morphologies of only spiral and irregular star-forming galaxies in the local universe. Therefore, according to these arguments, there 
could still be real differences in the spike and non-spike morphology distributions even though we do not detect significant evidence of 
differentiation with our analysis.  
\newline\indent
The fourth point is that, since the galaxies must be young (as indicated by their high redshift and stellar ages, typically $\sim 100$ Myr $-$ 
1 Gyr ), they may not have had time to settle into dynamical equilibrium.  Studies of galaxies forming by monolithic collapse 
\citep{katz1991,katz1992,immeli2004} or by mergers \citep{robertson2004} demonstrate that the timescales for galaxies to reach a recognizable 
disk or bulge morphology are of order Gyrs.  Therefore, morphology may be decoupled from star formation history at high redshift.  We explore 
this connection in the following section.

%\clearpage
%%%Environment

%\clearpage
\begin{deluxetable}{lcccc}
\tablewidth{0pc}
\footnotesize
\tablecaption{KS Test Comparison of Morphology and Environment} 
\tablehead{
\colhead{Pixel Selection Scheme} &\colhead{Size} &\colhead{Gini} &\colhead{Multiplicity} &\colhead{Flux}
}

\startdata
%G1 T2
\textbf{Fiducial} &0.78\tablenotemark{a} &0.61\tablenotemark{a} &0.18\tablenotemark{a} &0.69\tablenotemark{a} \\
%D T5 30, aka, David's non-z scheme
Scheme A &0.91 &0.73 &0.59 &0.88 \\
%Dz z=3.4 T3 30, aka, David's scheme
Scheme B &0.27 &0.33 &0.83 &0.44 \\
%G1 T2 30
Scheme C &0.68 &0.58 &0.71 &0.54
\enddata

\tablenotetext{a}{The KS probability that the spike and non-spike morphologies are drawn from the same distribution.}
\end{deluxetable}

\section{Morphology and Physical Properties of Galaxies}\label{sec:phys}
In order to examine morphology in the context of the star formation history and physical properties of galaxies, we use properties derived 
from the stellar population synthesis modeling of the broadband $U_n G \mathcal{R} J K_s +$\emph{Spitzer}/IRAC photometry performed by 
\citet{shapley2005a} and \citet{erb2006b}.  These parameters include stellar mass, age, extinction, and star formation rate. In the case of 
\citet{erb2006b}, $K_s$ band magnitudes are corrected for the H$\alpha$ emission line.  Stellar population parameters were only estimated for 
galaxies with both spectroscopic redshifts and near-IR photometry.  The criterion on the redshifts was imposed because simultaneous fitting of 
photometric redshifts and star formation history leads to large uncertainties in both the photometric redshifts (typically $|z_{spec} - 
z_{phot}| = 0.3$) and stellar population model parameters \citep{shapley2005a,vandokkum2006}.  It is necessary to have near-IR photometry 
because near-IR bands lie on the red side of the age-sensitive 4000$\mbox{\AA}$ and  Balmer break features, and hence, are crucial in 
separating young from old stellar populations.  There is a total of 60 BX/MD objects on the ACS pointings that satisfy both observational 
requirements, including 22 spike galaxies. 

\subsection{Methods: Stellar Population Synthesis Models and Statistical Tests}
The details of population synthesis modeling are described in \citet{shapley2005a} and \citet{erb2006b}.  Briefly, theoretical SEDs are 
derived assuming solar metallicity and a \citet{chabrier2003} initial mass function \citep{bc2003}.  Starlight is attenuated with the 
\citet{calzetti2000} dust attenuation model, which can be parameterized in terms of the extinction $E(B-V)$.   The star formation history is 
assumed to be a single burst $\tau-$model, such that the star formation rate $SFR(t_{sf}) = SFR_0 e^{-t_{sf}/\tau}$, where $SFR_0$ is the star 
formation rate at the beginning of the star-forming epoch, $t_{sf}$ is the time since the beginning of star formation (i.e., it is the age of 
the stellar population), and $\tau$ parameterizes the length of the epoch of star formation.  Constant star formation is equivalent to the 
statement that $\tau = \infty$.  The $\chi^2$ for each model is computed by taking the difference between the broadband photometry and the 
model SED.
\newline\indent
There are several items to note for our analysis below.  First, each model is parameterized by stellar age, stellar mass, star formation rate 
(SFR), extinction $E(B-V)$, and $\tau$.  Secondly, given the degeneracy between dust extinction and the age of the stellar population (or more 
generally, the star formation history), the $\chi^2$ distribution is fairly broad for most galaxies.  Fits with a constant star formation rate 
($\tau = \infty$) are generally almost as good as the best fit model.  This means that,in general, stellar age, $E(B-V)$, SFR, and $\tau$ are 
loosely constrained.  However, many authors \citep{papovich2001,shapley2001,shapley2005a} have shown that the assembled stellar mass is more 
robustly constrained.  Below, we will consider the model parameters of both the best fit stellar population synthesis model and the best fit 
model assuming a constant star formation rate.
%\newline\indent

\begin{deluxetable*}{llcccccc} %cc}
%\rotate
\label{tab:phys}
\tablewidth{0pc}
\footnotesize
\tablecaption{Correlations of morphological and physical parameters for BX/MD objects}

\tablehead{
	&	&\multicolumn{2}{c}{Size} &\multicolumn{2}{c}{Gini} &\multicolumn{2}{c}{Multiplicity} \\ %&\multicolumn{2}{c}{Flux} \\
\colhead{} &\colhead{} &\colhead{KS\tablenotemark{a}} &\colhead{$\mathrm{N}_\sigma$\tablenotemark{b}} &\colhead{KS } &\colhead{$\mathrm{N}_\sigma$} &\colhead{KS} &\colhead{$\mathrm{N}_\sigma$} %&\colhead{KS} &\colhead{$\mathrm{N}_\sigma$}
}
\startdata
Constant SFR  &Age	&0.059142 	&1.20		&0.275269	&1.92	&0.275269	&$-0.75$\\%	&$0.000725$	&2.69 \\
	&Mass		&0.059142	&$-1.48$	&0.023213	&1.87	&0.275269	&$-0.96$\\%	&0.770953	&0.48 \\
	&E(B-V)		&0.965485	&$-0.49$	&0.023213	&1.19	&0.059142	&1.03\\%		&0.275269	&0.70 \\
	&SFR		&0.275269	&$-1.53$	&0.770953	&$-0.90$	&0.134947	&1.09\\%	&0.023213	&$-1.91$ \\
Best Fit &Age		&0.134947	&1.41		&0.497342	&1.74	&0.134947	&$-1.08$\\%	&$0.000725$	&2.88 \\
	&Mass		&0.023213	&$-1.69$	&0.023213	&1.66	&0.134947	&$-1.39$\\%	&0.965485	&0.31 \\
	&E(B-V) 	&0.770953	&0.79		&0.134947	&1.01	&$0.002571$	&1.97\\%		&0.275269	&1.41 \\
	&SFR		&0.965485	&1.09		&0.965485	&$-0.67$	&0.275269	&1.20\\%	&0.965485	&0.34 \\
	&	&	&	&	&	&	&	\\%&	&	\\
UV Absolute Magnitude	&	&$0.002929$	&3.73	&$<0.000001$	&5.29	&$0.490256$	&$-0.33$%	&$<0.000001$	&7.50 \\

%Lyman-$\alpha$ EW\tablenotemark{c} &	&$0.007054$ &3.05 &0.348120	&$-1.32$	&0.162466	&0.50	&0.065669	&1.77
\enddata

\tablenotetext{a}{The KS probability that the morphological parameters in the lowest physical parameter bin are drawn from the same distribution as those in the highest physical parameter bin.}
\tablenotetext{b}{Number of standard deviations from the null hypothesis for the Spearman correlation test.}
%\tablenotetext{c}{There were 50 objects with Ly-$\alpha$ equivalent width measurements.}

\end{deluxetable*}

%\clearpage
\begin{figure*}
\epsscale{1.15}
\plottwo{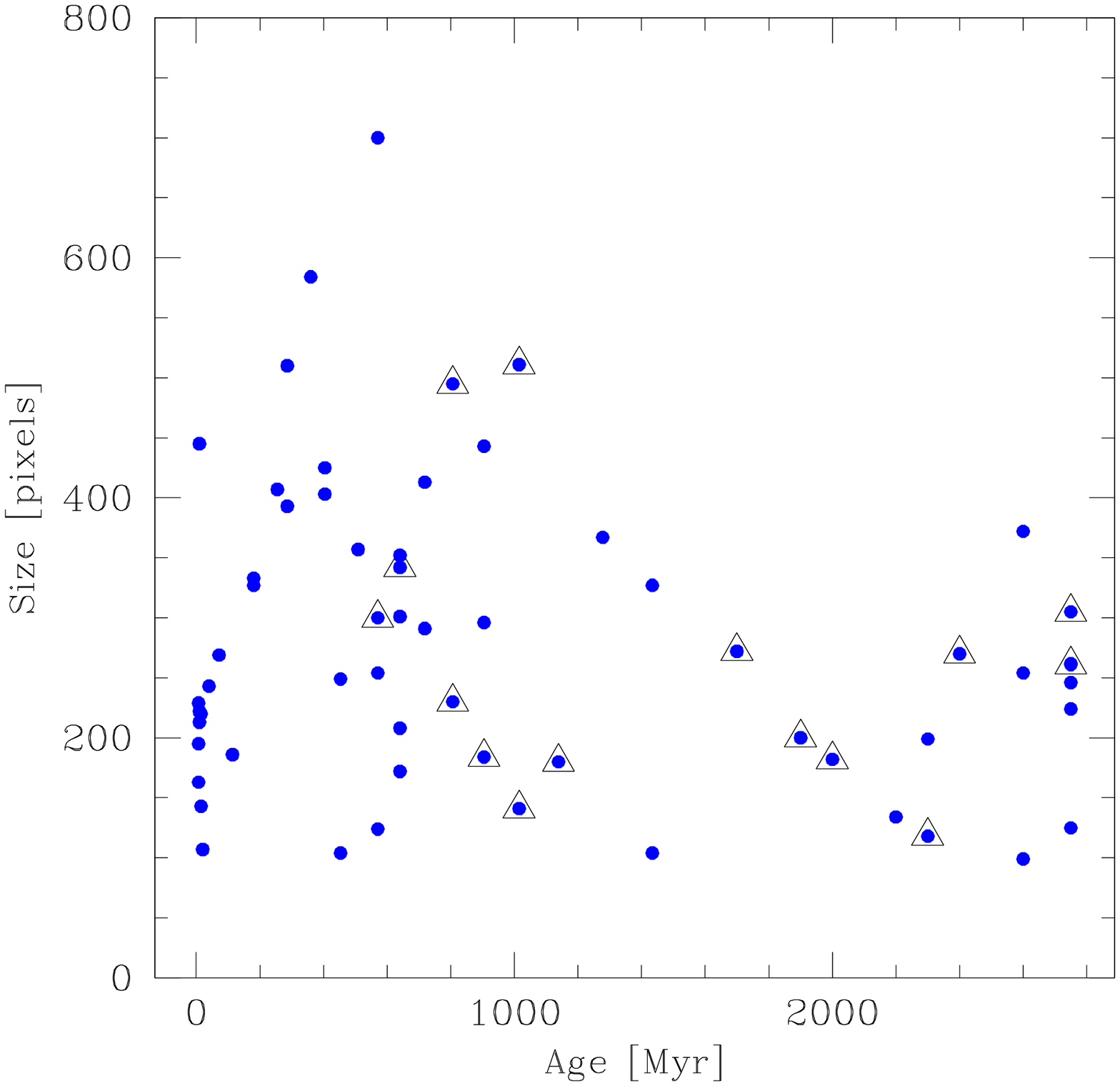}{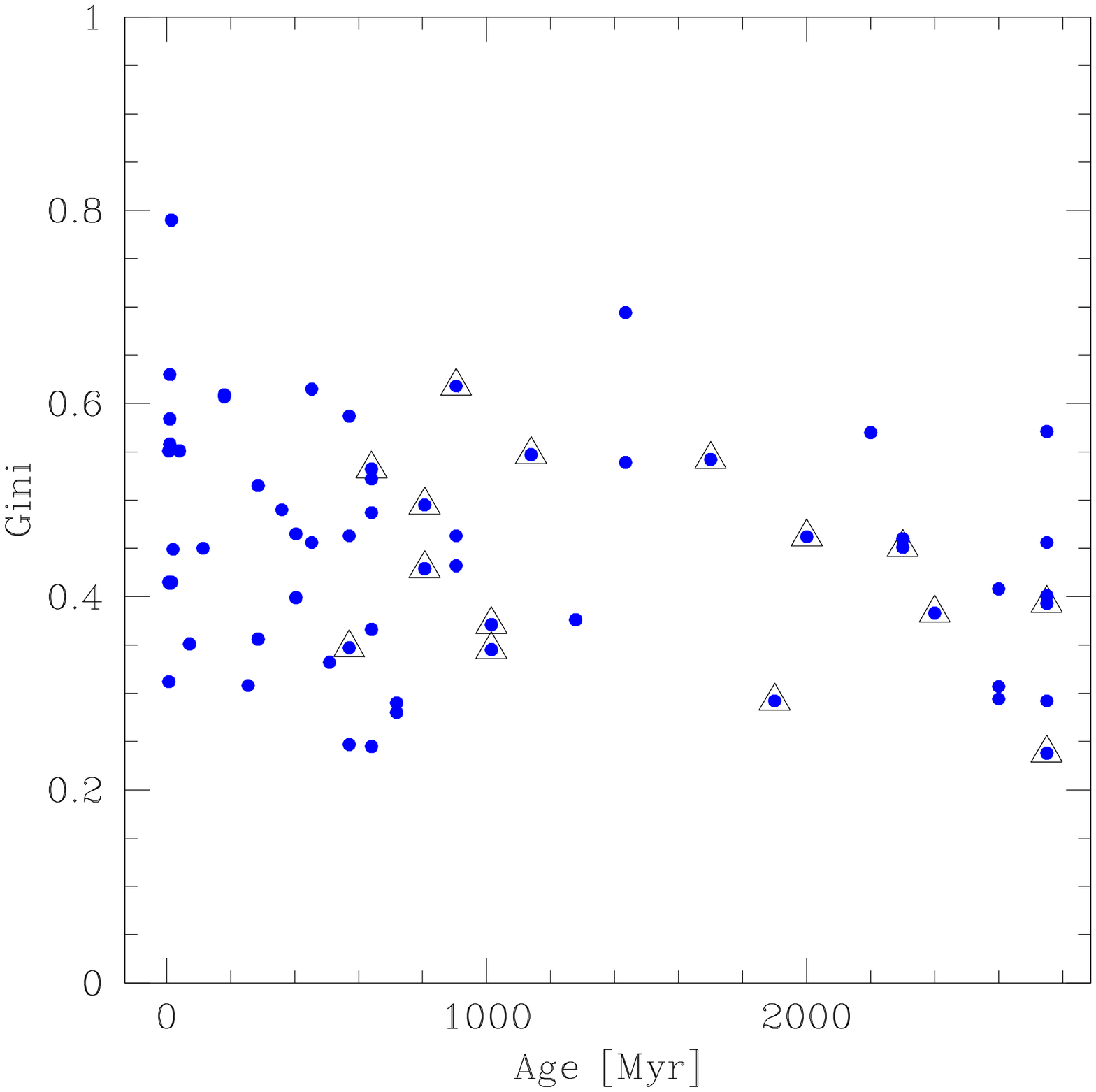}
\caption{The correlations between stellar age (derived from the best-fit stellar population model with constant star formation rate) and morphological parameters.  Open triangles mark spike galaxies.  Left: Size as a function of age.  Right: Gini as a function of age.}
\label{fig:age}
\end{figure*}

In addition to examining the relationships between star formation history and morphology, we also investigate any connections between 
morphology and rest-frame UV luminosity.  To determine the rest-frame UV luminosity, we use the \emph{G}-band magnitude (centered at 4780 
$\mbox{\AA}$) and neglect K-corrections and extinction corrections.  Therefore, the UV luminosity is centered at wavelengths dependent upon 
the redshift of the object; the \emph{G} band corresponds to a rest-frame luminosity centered at $1450 \mbox{\AA}$ for the spike mean redshift 
of $z=2.300$.  For the range of redshifts including 80\% of the BX/MD spectroscopic sample ($z=1.9-2.8$), the \emph{G} band probes rest-frame 
luminosities centered at $1650 - 1260 \mbox{\AA}$.
\newline\indent
We used two different statistical measures to evaluate relationships between morphological and physical properties.  For the first test, we 
sorted galaxies by the physical parameter in question.  We then divided the galaxies into three bins, and performed one-dimensional KS tests 
on the morphological distributions in the highest and lowest bins (as determined by the physical parameter).  This test allows us to explore 
if, for example, the most massive third of galaxies have different morphology distributions than the least massive third of galaxies.  
Secondly, we performed Spearman rank correlation tests for each pair of morphological and physical parameter distributions.  The results of 
our analysis are shown in Table 3.  We quantify the results of the KS tests in terms of the probability that the morphological distributions 
in the uppermost and lowermost physical parameter bins are drawn from the same distribution.  We quantify the results of the Spearman test in 
terms of the number of standard deviations from the expectation value of the hypothesis that the data are uncorrelated.

%\clearpage
\begin{figure*}
\epsscale{1.15}
\plottwo{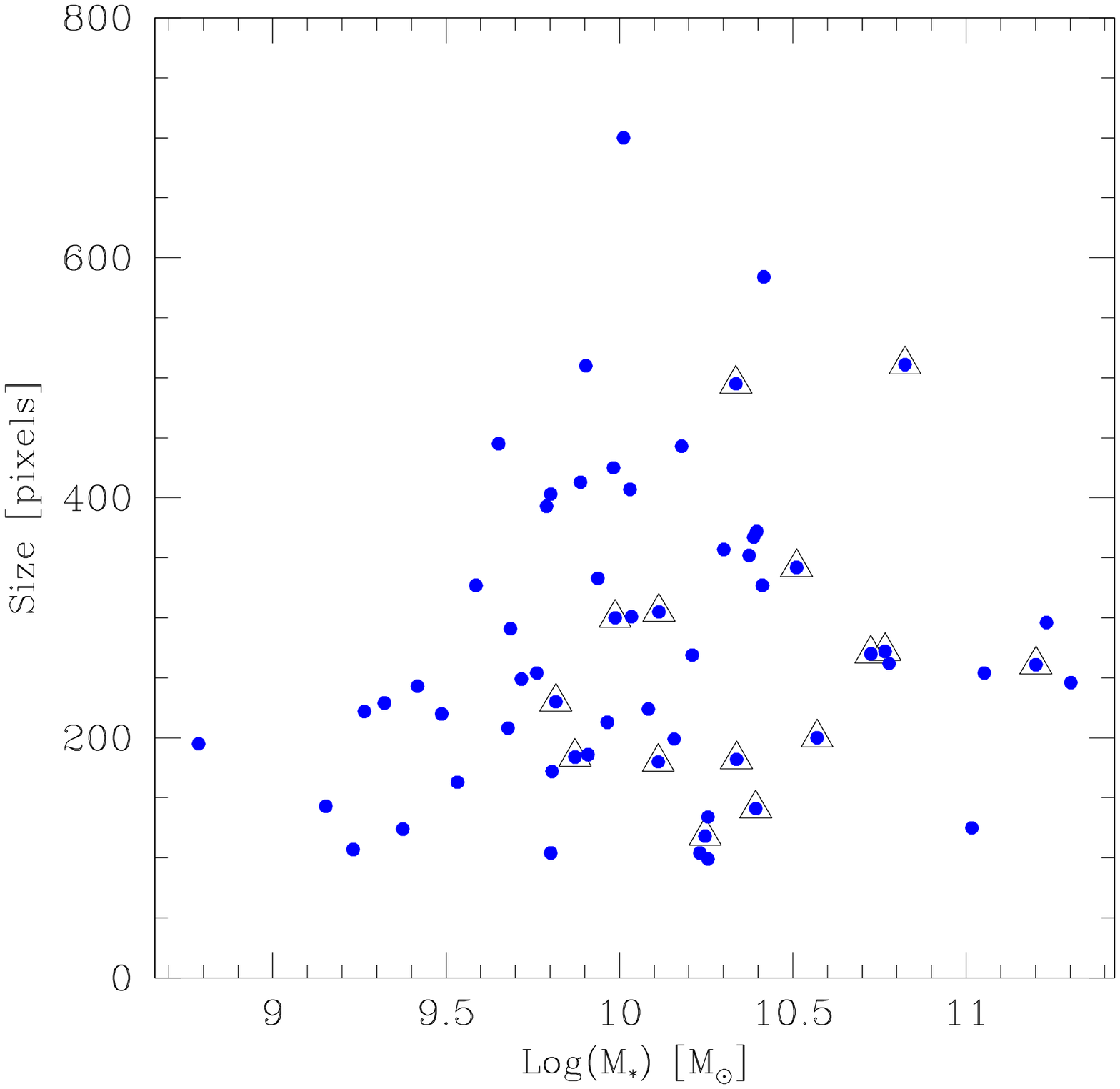}{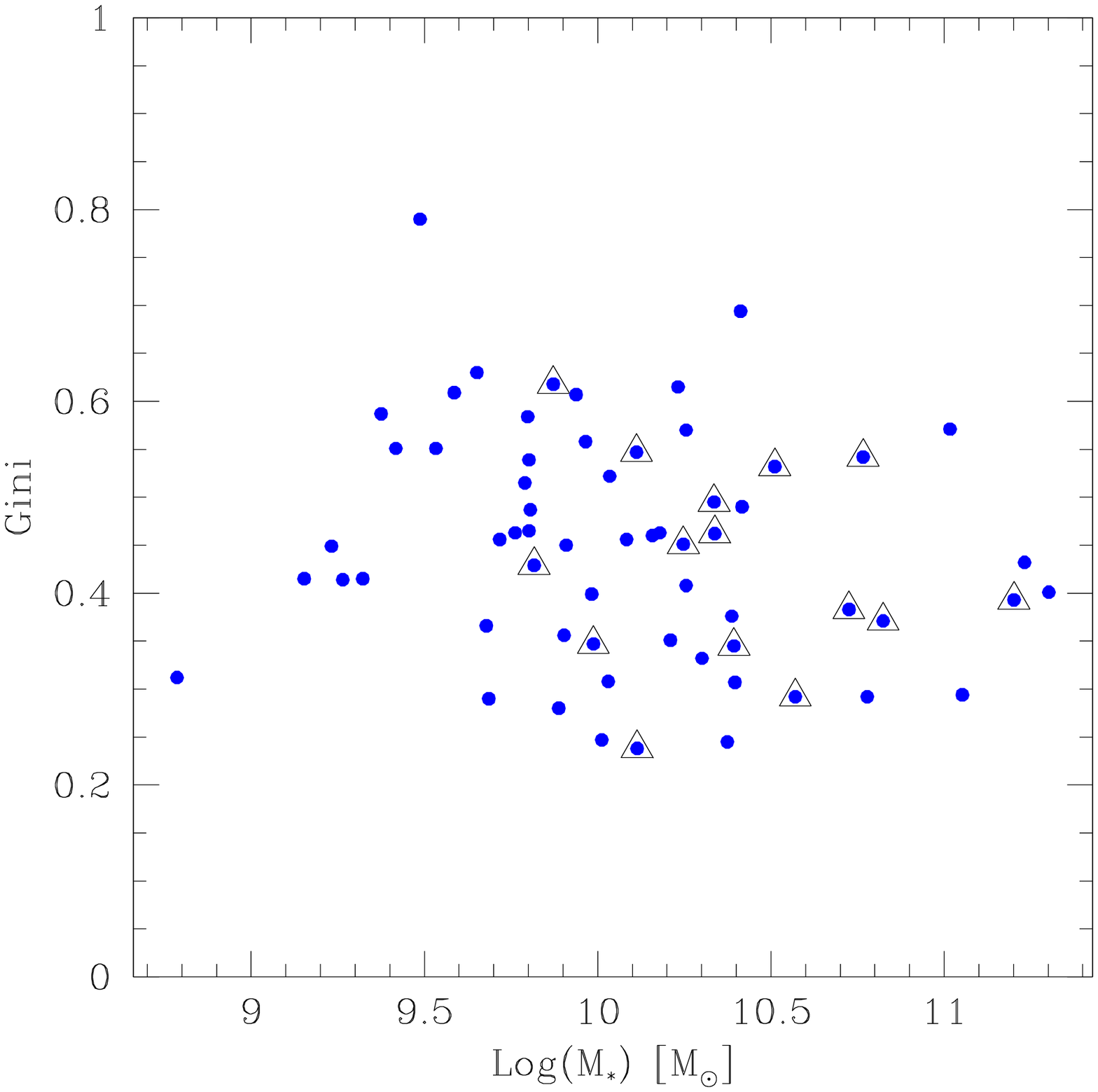}
\caption{The correlations between stellar mass (derived from the best-fit stellar population model with constant star formation rate) and  morphological parameters.  Open triangles mark spike galaxies.  Left: Size as s function of mass.  Right: Gini as a function of mass.}
\label{fig:mass}
\end{figure*}

\subsection{Morphology and Stellar Populations}
The main result from our statistical tests is that there are very few strong correlations between physical properties and morphology.  First, 
we consider stellar mass and age, which in \citet{steidel2005} showed significant differences in distribution as a function of environment.  
For both the best fit $\tau$ and constant star formation models, age and mass are only weakly correlated with galaxy size (at the $\sim 
1.5\sigma$ level) and gini (at slightly less than $2\sigma$).  In Figures \ref{fig:age} and \ref{fig:mass}, we show scatter plots in the 
age-size, age-gini, mass-size, and mass-gini planes to demonstrate how tenuous the correlations are.  Moreover, our statistical tests 
demonstrate that multiplicity is not well correlated with either stellar age or stellar mass.  The fact that morphology correlates only weakly 
with stellar age and stellar mass reconciles the finding that while age and mass are strongly correlated with environment \citep{steidel2005}, 
morphology is not.
\newline\indent
Next, we comment on the extinction and star formation rate results.  As demonstrated in Table 3, the correlation strength between morphology 
and $E(B-V)$ or SFR are generally weak.  Moreover, the deviations from the null hypothesis that morphology and $E(B-V)$ or the SFR are 
uncorrelated are not always in the same direction, depending on whether the physical quantities are estimated from the best fit $\tau$-model 
or constant star formation model.   The lack of consistency is a statement about the inability of the stellar population synthesis models to 
strongly constrain the values of $E(B-V)$ and SFR.  Given the large error bars on these parameters, it is also unlikely that the KS and 
Spearman tests yield believable information on the strength of correlation between the $E(B-V)$ and morphology, and SFR and morphology.  It 
should be noted that our results are in contrast with those of \citet{law2007}, who find a significant correlation between $E(B-V)$ and both 
size and gini.  \citet{law2007} estimated $E(B-V)$ using best-fit $\tau$-models.  Given the uncertainty in estimating the extinction with SED 
fitting, especially when $\tau$ is allowed to vary, it is not so surprising that the correlation tests for these relatively small samples 
yield different results.

\subsection{Morphology and UV Luminosity}
\begin{figure*}
\epsscale{1.15}
\plottwo{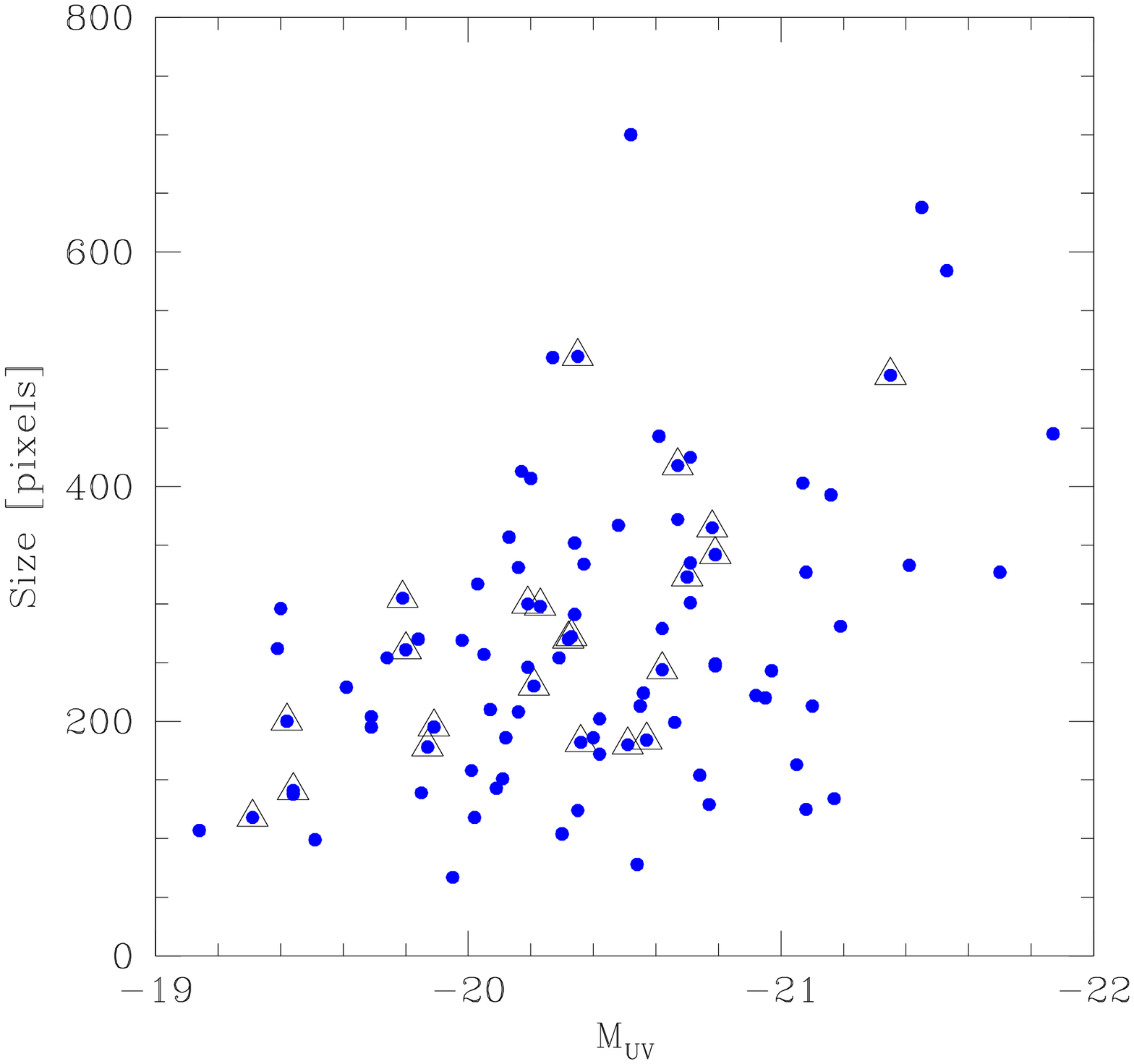}{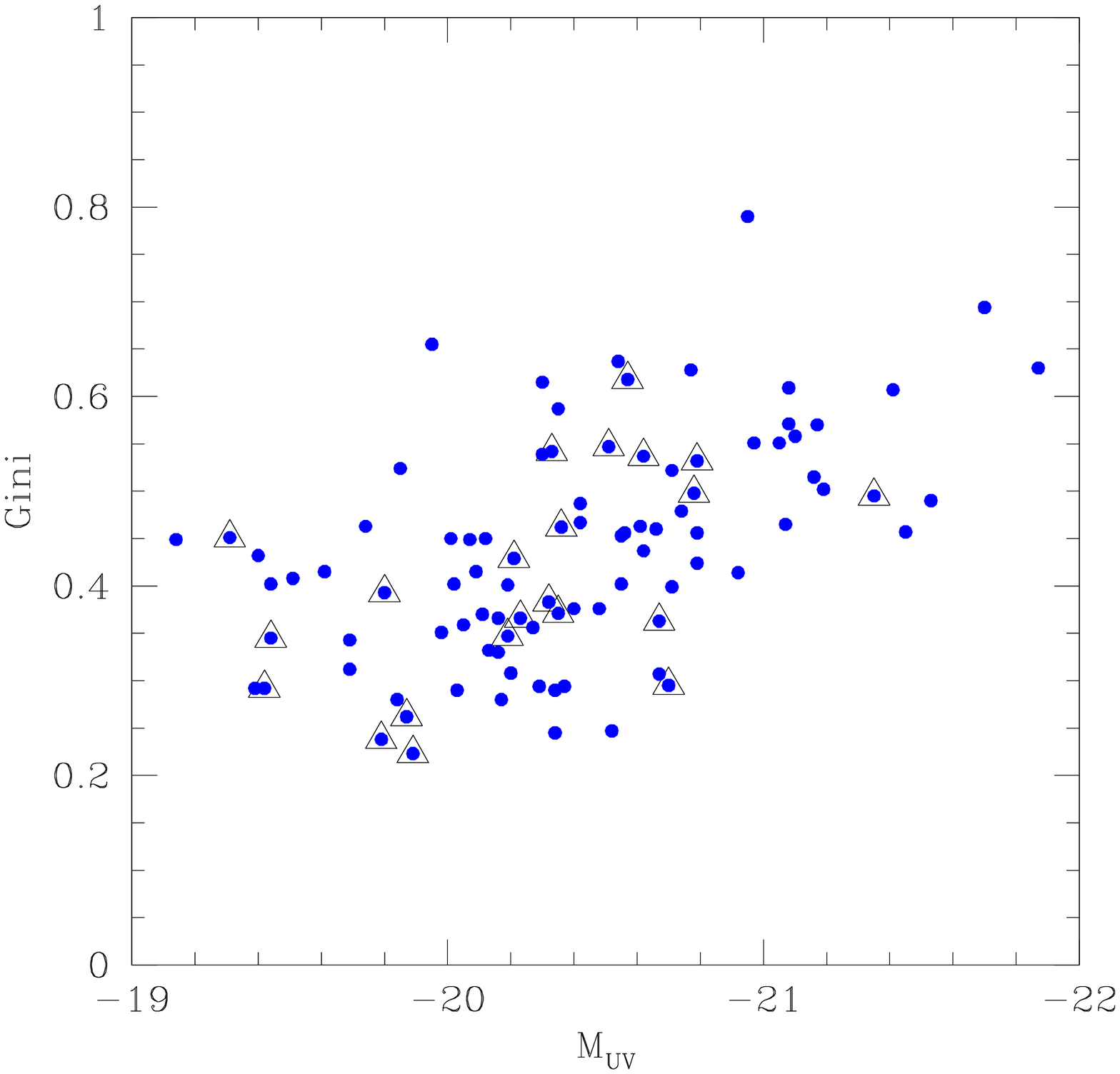}
\caption{The correlation between UV absolute magnitude and galaxy size (left) and gini (right) for the spectroscopic BX/MD sample.  Open triangles mark spike galaxies.}
\label{fig:m}
\end{figure*}

We find the strongest correlations between UV luminosity (or absolute magnitude) and the morphological parameters gini and size (Figure 
\ref{fig:m}, Table 3), while there is no evidence for a correlation between multiplicity and UV luminosity.  We perform several tests to 
determine if the observed correlations with UV luminosity are robust.  First, we perform Spearman rank correlation tests for a variety of 
pixel selection schemes, to ascertain that the correlations do not depend on the pixel selection method.  As demonstrated in Table 4, the 
correlations appear to be robust to pixel selection method.

%\clearpage

%\newline\indent
Secondly, we consider that surface brightness dependent pixel selection schemes might bias the luminosity-size and luminosity-gini 
relationships.  The fundamental reason why one ought to be concerned about surface brightness dependent assignments to galaxies has its origin 
in our definitions of gini and galaxy size.  Gini is a parameter that quantifies the flux distribution function (i.e., the number of pixels of 
given flux).  However, given that our instrument does not have infinite sensitivity, we calculate gini with only those pixels whose surface 
brightness is above a certain threshold.  To make an analogy with economics (the field which gave birth to the gini coefficient), applying a 
surface brightness threshold cut would be the economic equivalent of including only those whose income is above the poverty line in 
calculating the distribution of wealth in a population.  In economic literature, such a restriction on the distribution is called a 
``truncation bias.''  Since we also define size in a surface brightness-dependent way, as opposed to the traditional half-light radius, size 
is also subject to truncation bias.  
%\newline\indent

%\clearpage

\begin{deluxetable}{lcccc}
\label{tab:lumalt}
\tablewidth{0pc}
\footnotesize
\tablecaption{Correlations with UV absolute magnitude} 

\tablehead{
\colhead{Scheme} &\colhead{Size} &\colhead{Gini} &\colhead{Multiplicity} &\colhead{Flux} 
}

\startdata
%G1 T2
\textbf{Fiducial}	&3.73\tablenotemark{a}	&5.29\tablenotemark{a}	&$-0.33$\tablenotemark{a}	&7.50\tablenotemark{a} \\
%D T5 30
A 	&5.40	&5.17	&$-1.45$	&7.17\\
%Dz z=3.4 T3 30
B 	&6.39	&5.49	&$-1.07$	&7.27\\
C 	&4.05	&5.24	&$-0.23$	&7.26\\
\enddata

\tablenotetext{a}{Number of standard deviations from the null hypothesis that the given parameter is uncorrelated with UV absolute magnitude, as determined by the Spearman rank correlation test.}
\end{deluxetable}

%\clearpage
%%%Flux distribution functions
\begin{figure}
\plotone{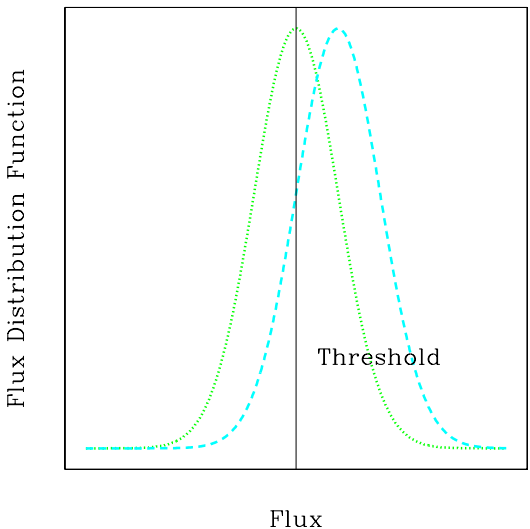}
\caption{Schematic flux distribution functions for two galaxies of different luminosity but with the same form of the distribution function.}
\label{fig:trunc}
\end{figure}

%\clearpage
%%% Gini image with flux distributions
\begin{figure*}
\epsscale{1.15}
\plottwo{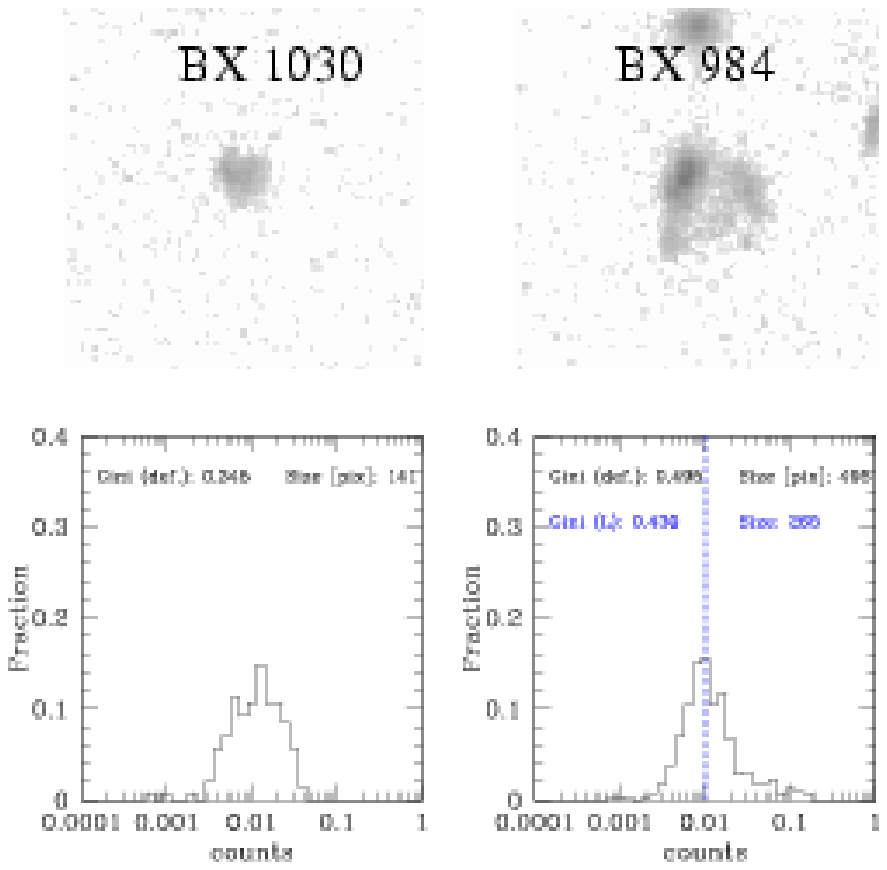}{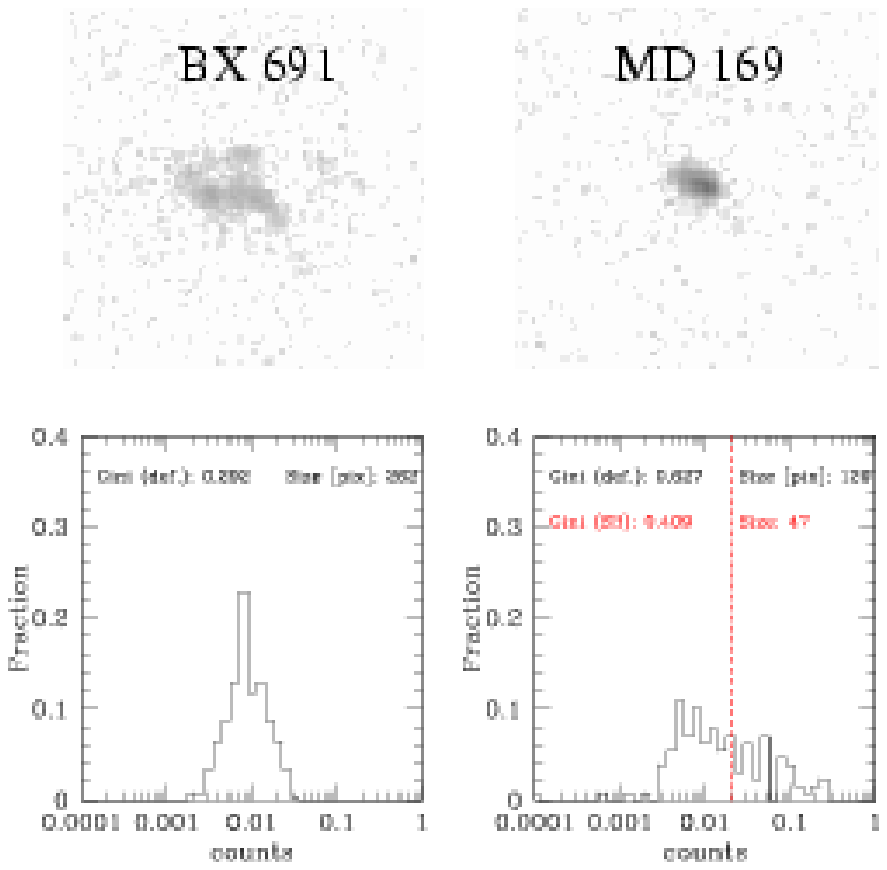}
\caption{Upper panels: \emph{HST}/ACS images of (\emph{left} to \emph{right}) the low luminosity
galaxy BX 1030, the high luminosity galaxy BX 984, the low surface brightness galaxy BX 691, and the
high surface brightness galaxy MD 169.  Each galaxy has typical morphology for the galaxies within
the sample (i.e., BX 1030 has typical gini and size of the ten lowest luminosity galaxies).  The
images are stretched on a logarithmic scale to match the flux distributions below.  Lower
panels: Flux distribution functions for the galaxy above.  We label the gini and size determined with
the fiducial pixel selection scheme as ``default.''  In the leftmost two panels, we consider the
effects of the surface brightness threshold on the gini-luminosity and size-luminosity correlations.
The dashed line indicates the surface brightness threshold scaled from its default value as the ratio
of the mean surface brightness of the 10 most luminous galaxies to the 10 least
luminous galaxies.  Both gini and size are recalculated for the typical high luminosity galaxy BX 984
using the scaled surface brightness threshold, and are labeled ``L.''  In the rightmost two panels,
we consider the effects of the surface brightness threshold on the gini-surface brightness and
size-surface brightness correlations.  The dashed line indicates the surface brightness threshold
scaled from its default value as the ratio of the mean surface brightness of the 10 highest surface
brightness galaxies to the 10 lowest surface brightness galaxies.  Both gini and size are
recalculated for the typical high surface brightness galaxy MD 169  using the scaled surface
brightness threshold, and are labeled ``SB.''}
\label{fig:giniflux}
\end{figure*}

\begin{figure*}
\epsscale{1.15}
\plottwo{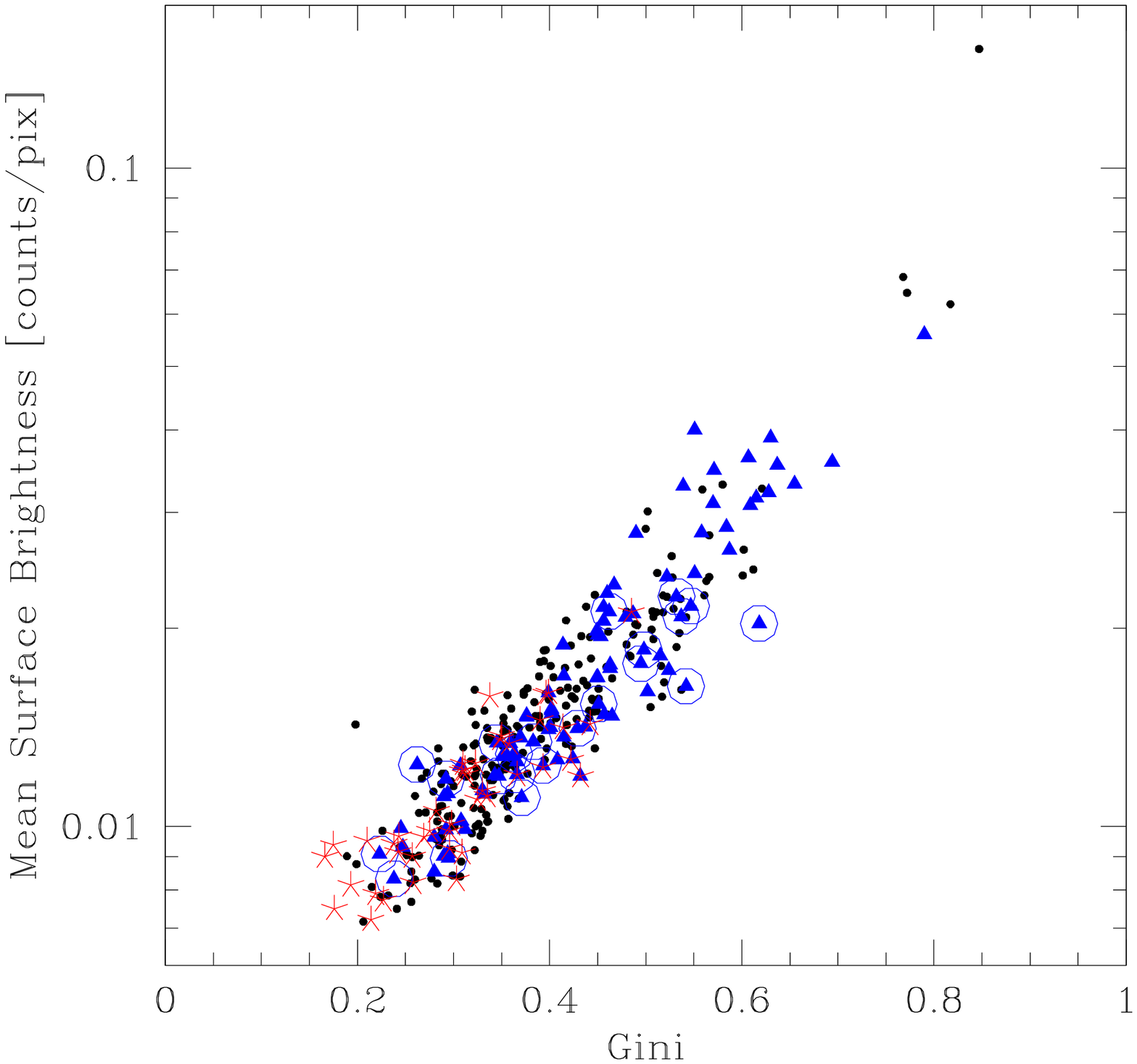}{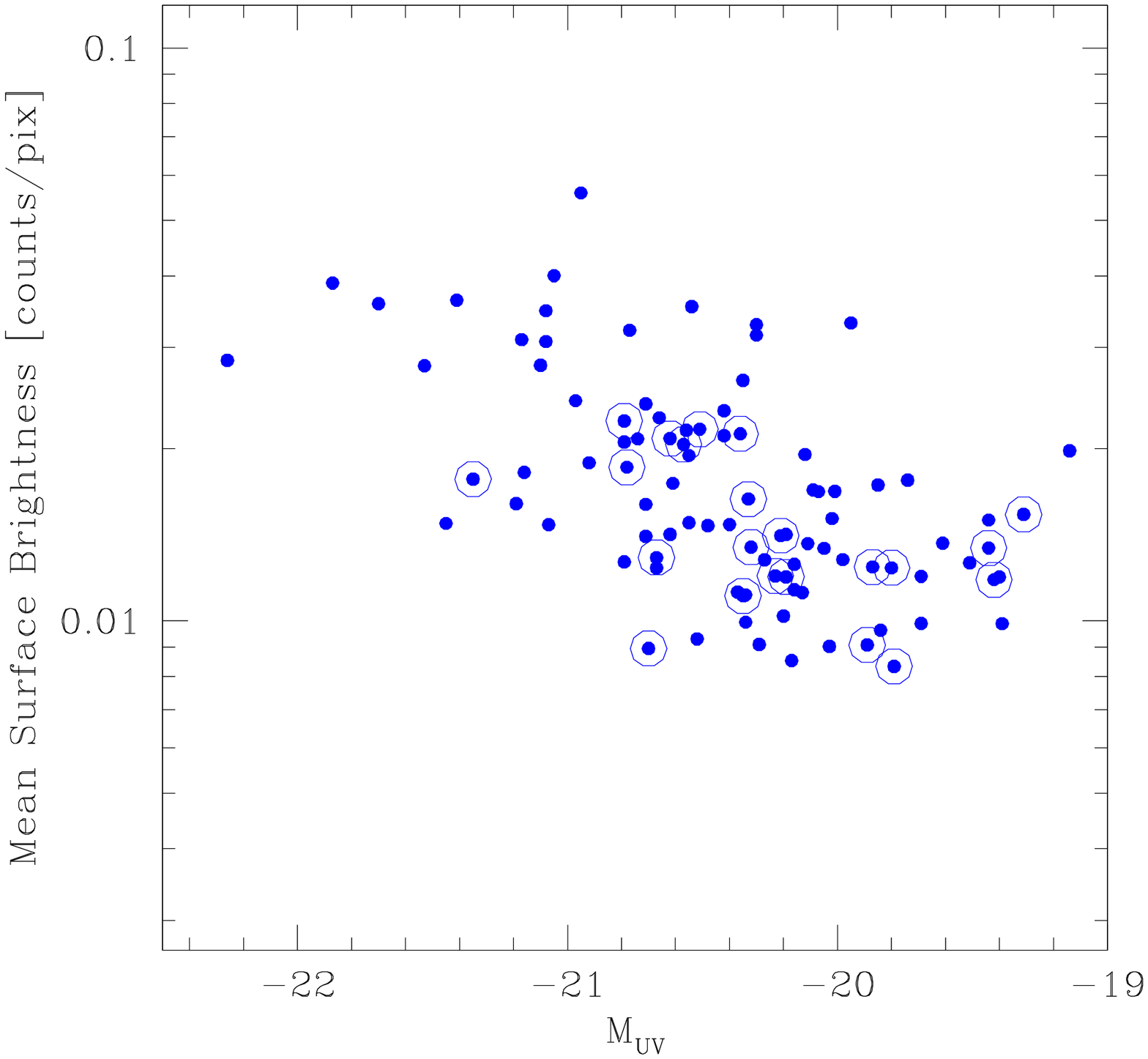}
\caption{Surface brightness as a function of gini (left); surface brightness as a function of UV absolute 
magnitude (right).  The small black dots denote members of the photometric BX/MD sample.  Spectroscopically 
confirmed BX/MD galaxies are marked with larger blue triangles, with members of the spike outlined with blue 
circles.  Red asterisks mark the DRG sample.}
\label{fig:sb}
\end{figure*}

It is important to consider what a constant surface brightness threshold means when considering galaxies with a range of luminosities.  Using 
an economic analogy, we want to consider what gini means if we were calculating gini for two countries (e.g., the United States and Mexico) 
with different GDPs, using incomes measured above a threshold fixed to the poverty line of one of the countries (e.g., fixing the threshold 
for both countries at the US poverty line).  Getting back to gini's astrophysical context, one can consider this question in a more visual way 
by examining the flux distribution function of Figure \ref{fig:trunc}.  If the distribution of pixels was such that the surface brightness 
threshold cut near the peak of the flux distribution function, one would expect the morphological parameters to deviate significantly from the 
values they would have had if the surface brightness threshold were lower.  If two galaxies at the same redshift have flux distributions with 
identical shape, but different normalization (i.e., the galaxies have different luminosity), one would expect that much less of the fainter 
galaxy's flux distribution would meet the surface brightness criterion than the brighter galaxy's.  In particular, one would expect the the 
size of the fainter galaxy to be artificially small because fewer pixels would be associated with the galaxy.  Additionally, the gini of the 
faint galaxy would be artificially small because the dynamic range of the fluxes in the pixels would be suppressed.  In an example such as 
this, the ideal thing to do would be to scale the surface brightness threshold so that the threshold would apply for the same point on the 
flux distribution function curve, relative to the mean surface brightness, for each galaxy.  This is the economic equivalent of scaling the 
threshold for calculating gini by the mean income per person in each country.
\newline\indent
In order to estimate how much gini and size are affected by surface brightness thresholds, we focus on 
the
ends of the luminosity distribution.  The ten lowest luminosity galaxies have a mean surface
brightness of $\langle SB\rangle_{low} = 0.015$ counts/pix/sec, as calculated using the fiducial 
pixel  
selection method, mean gini $\bar{G} = 0.332 \pm 0.022$\footnote{The errors listed are the standard
deviation of the mean, not the sample standard deviation.}, and mean size $\bar{I} = 152\pm 18$
pixels.  The
ten highest luminosity galaxies had a mean surface brightness of $\langle SB\rangle_{high} = 0.03$
counts/pix/sec, mean gini $0.527\pm0.025$\, and mean size $\bar{I} = 447 \pm 92$ pixels, as
calculated using
the fiducial pixel selection method.  Representative galaxies (low luminosity: BX 1030, high 
luminosity: BX
984) for each sample are shown in Figure \ref{fig:giniflux}.  Since the mean surface brightness of
the highest
luminosity galaxies is a factor of two greater than the mean surface brightness, we scale the surface
brightness threshold for the high luminosity galaxies by that same factor of two in order to estimate 
the
effects of truncation bias on the gini-luminosity and size-luminosity relations.  When we select 
pixels for
the high luminosity sample using this higher surface brightness threshold, we find that the mean gini
reduces
to $\bar{G} = 0.467 \pm 0.024$, and the mean size to $\bar{I} = 292 \pm 65$ pixels (see the lower 
panels of
Figure \ref{fig:giniflux} for an example of how the surface brightness threshold affects size and
gini).  Both
the mean gini and
size of the high luminosity sample, as calculated with the scaled surface brightness threshold, are $>
1\sigma$ from the means of the low luminosity sample (calculated using the default pixel selection
method).
This suggests that the correlation between size or gini and luminosity is real, and not the result of
truncation bias.  However, we caution that the
mean surface brightnesses were calculated using only pixels that were above the surface brightness
threshold, and hence the calculated mean
surface brightnesses are subject to truncation bias.
\newline\indent
We note that there is also a very tight correlation between the gini coefficient and mean surface  
brightness (Figure \ref{fig:sb}), while there is only a weak ($2 \sigma$) anti-correlation between
size and
surface brightness.  In order
to determine if low surface brightness galaxies have artificially low ginis due to truncation bias,
we raise the surface brightness threshold of the highest surface
brightness galaxies by the ratio (a factor of four) of the mean surface brightness of the
ten highest surface brightness galaxies to
the mean surface brightness of the ten lowest surface brightness galaxies (Figure \ref{fig:giniflux}).
We then recalculate the ginis of the
highest surface brightness galaxies using pixels which meet the new surface brightness threshold,
just as we did with luminosity.  The mean gini of the high surface brightness sample (using the
fiducial pixel selection method) is
$0.599\pm 0.024$, and the mean gini of the low surface brightness sample is $0.227\pm
0.009$.  We find that mean gini of the highest
surface brightness galaxies, calculated with the increased surface brightness threshold, decreases
by $\sim 0.2$ to $0.444\pm 0.026$, which is not enough to destroy
any of the gini-surface brightness correlation.
%\newline\indent

%\clearpage

The above trends are consistent with the results
from simulations of our
morphological analysis procedure. As described in Section~\ref{sec:pix}, 
in these simulations
we added a series of artificial object images to the real ACS image. 
In this case, the artificial
images consisted of the pixel light distributions for objects
among the brightest in our sample, with the highest gini vales.
We then recovered
the corresponding morphological parameters for
simulated objects in relatively clean locations. Next,
we ran the same simulations after scaling down
the simulated object images by factors of 2 and 4
(roughly equivalent to lifting the surface brightness
threshold by the same amount). We found that gini and 
size for the scaled, fainter
object images are lower by amounts consistent
with what is outlined above.
\newline\indent
Closer examination of the flux distributions demonstrate (c.f., Figure \ref{fig:giniflux}) that the
flux distribution functions of low gini
and high gini objects have fundamentally different shapes.  In particular, the most luminous galaxies 
and
those galaxies with the highest surface brightness  
have most of the flux concentrated
in just a few pixels.  This means that both the gini coefficient and average surface brightness must
necessarily be quite high for high luminosity galaxies.
\newline\indent
We also consider the fact that we do not K-correct the UV luminosities to a single waveband.  Therefore, the UV luminosity we use in our 
analysis effectively samples the galactic SEDs at a redshift-dependent wavelength.  This could affect a correlation in the following sense. If 
there is a tilt to the typical SED, the \emph{G} band (from which we derive UV luminosities) will probe a different part of the SED as a 
function of redshift.  Therefore, lower redshift galaxies could appear systematically brighter or fainter than their higher redshift 
counterparts.  If the morphologies had a redshift dependence, this would imprint itself in the luminosity-morphology relations.  In order to 
see if underlying size-luminosity and gini-luminosity correlations exist if the UV luminosity samples the same part of the galaxy SED, we 
perform Spearman correlation tests on the spike galaxies alone.  The central wavelengths for the spike range from 1442 to 1455$\mbox{\AA}$, a 
range much narrower than that of the entire spectroscopic BX/MD sample.  Since the morphologies and luminosities of spike galaxies are not 
significantly different from those of field galaxies, we expect any morphology-related correlation present in the spike galaxies to also be 
present in the larger population.  We find that, for the fiducial pixel selection scheme, size and UV luminosity are still correlated at the 
$2.7\sigma$ level, and gini and UV luminosity are correlated at the $2.5\sigma$ level.  Therefore, the UV luminosity-size and UV 
luminosity-gini correlations appear to be real.
%\newline\indent
%Lastly, we consider correlations between the Lyman$-\alpha$ equivalent width and morphology.  Table 3 demonstrates a $3\sigma$ correlation between size and Lyman$-\alpha$ equivalent width, with no other correlations being more than $2\simga$ in significance.  Upon closer examination of the size-Lyman$-\alpha$ equivalent width plane, we find that there is very little scatter in size of galaxies which have high positive Lyman$-\alpha$ equivalent width, but significant scatter in the sizes of galaxies with negative equivalent width.  In general, it seems that galaxies with Lyman$-\alpha$ in emission are smaller than those with Lyman$-\alpha$ in absorption.

\subsection{Comparison with Recent Work}
%In summary, we find only a few significant correlations between morphology and the physical 
%properties and star formation histories of galaxies.  Interestingly, we find no significant 
%correlation between multiplicity and any physical parameter.  We find only weak 
%correlations (at the $1.5-2\sigma$ level) between stellar age and size, 
%stellar age and gini, stellar mass and size, and stellar mass and gini.  
%However, the strongest correlations are between UV luminosity and size, and UV 
%luminosity and gini.  These correlations appear to be robust to the several tests we 
%described above.  The size-luminosity correlation is not surprising; in the 
%local universe, too, brighter galaxies are larger than fainter ones \citep{shen2003}.  
%Strong correlations between luminosity and size have also been found in the UDF 
%sample of galaxies spanning the redshift range $0 \leq z \lesssim 1$ \citep{cameron2007}.  
%The gini-luminosity correlation can be understood in the context of the flux distribution.  
%Most of the luminosity in a high luminosity galaxy comes from one or a few small, 
%unusually bright regions.  This means that flux is very unequally distributed among pixels, 
%and so the gini must necessarily also be high.  For low luminosity galaxies, there is an 
%absence of the very bright regions, and so gini must necessarily be lower.  
%\newline\indent
It is useful to compare our findings to those of \citet{law2007}, whose sample of 
spectroscopically confirmed $z\sim 2$ BX/MD galaxies in the GOODS-N has a redshift 
distribution that significantly overlaps with our sample.  We find that, 
when we calculate the morphological parameters of our spectroscopic 
BX/MD sample in the same way as \citet[][``Scheme B'']{law2007}, the 
mean values of size, gini, and multiplicity are almost identical to 
those of the $z\sim 2$ sample in the GOODS-N field. 
\newline\indent
In general, our results are consistent with those of \citet{law2007}.  
In particular, \citet{law2007} also find a strong correlation between 
gini and luminosity, and size and luminosity.  Our results on the 
strength of the correlation between morphology and stellar age and 
mass are mildly inconsistent with \citet{law2007}, who find no correlation 
between stellar mass and age and morphology.  We find weak evidence for 
correlations between stellar age and mass and morphology; however, the 
correlations only marginally significant at the $1.5-2\sigma$ level.  The main 
difference between the two analyses lies in the relationship between extinction and
morphology.  \citet{law2007} find a 3.3$\sigma$ correlation between extinction and 
size, and a 2.4$\sigma$ correlation between extinction and gini.  We do not 
find a correlation between extinction and any morphological parameter.  
However, the method of estimating extinction with SED fitting is subject to 
large errors.  Therefore, the lack of consistency in the strength of the 
extinction-morphology relationship in the two analyses is not meaningful.

%\clearpage

\begin{deluxetable*}{lcccccc}
%\rotate
\tablewidth{0pc}
\label{tab:samp2}
\footnotesize
\tablecaption{Mean and 1-$\sigma$ Standard Errors of Morphological Parameters for Each Galaxy Sample}

\tablehead{
	&\multicolumn{2}{c}{All}	&\multicolumn{2}{c}{$\langle \mathrm{SB} \rangle < 0.01$ count/pix/sec}	&\multicolumn{2}{c}{0.01 $\le \langle \mathrm{SB} \rangle < 0.02$ count/pix/sec} \\
\colhead{Parameter} &\colhead{BX/MD} &\colhead{DRG} &\colhead{BX/MD}	&\colhead{DRG}  &\colhead{BX/MD}	&\colhead{DRG}
}

\startdata
Number of Galaxies	&282	&43	&39	&19	&185	&23	\\
Size[pix]	&$251\pm 10$	&$215\pm 31$	&$272\pm 23$	&$158\pm31$	&$252\pm12$	&$269\pm 49$\\
Gini		&$0.402\pm 0.006$	&$0.308\pm 0.012$	&$0.269\pm0.006$	&$0.240\pm0.010$	&$0.381\pm0.005$	&$0.356\pm 0.010$\\
Multiplicity	&$6.10\pm 0.29$	&$9.36\pm 1.05$ &$10.44\pm0.86$	&$12.10\pm1.81$	&$6.17\pm0.32$	&$7.49\pm 1.06$\\
Flux[counts/sec]	&$4.12\pm 0.23$	&$2.51\pm 0.41$	&$2.46\pm0.21$	&$1.41\pm0.29$	&$3.54\pm0.16$	&$3.49\pm 0.67$\\	
$\langle \mathrm{SB} \rangle$ [$10^{-2}$ counts/pix/sec]	&$1.64\pm 0.11$ &$1.17 \pm 0.36$	&$0.90\pm 0.11$	&$0.89\pm 0.25$	&$1.40 \pm 0.09$	&$1.30\pm 0.34$
\enddata

\end{deluxetable*}

%\newline\indent
The most interesting result from this part of our analysis is that there is no strong 
correlation between the physical properties or star formation history and galaxy morphology.  
This is in stark contrast to what is observed at the present epoch, where star 
formation history and morphology are closely tied \citep{kauffmann2004},
though consistent with the results of \citet{law2007}.

\section{Morphologies of Optical- and IR-Selected Galaxies}
\begin{figure*}
\plotone{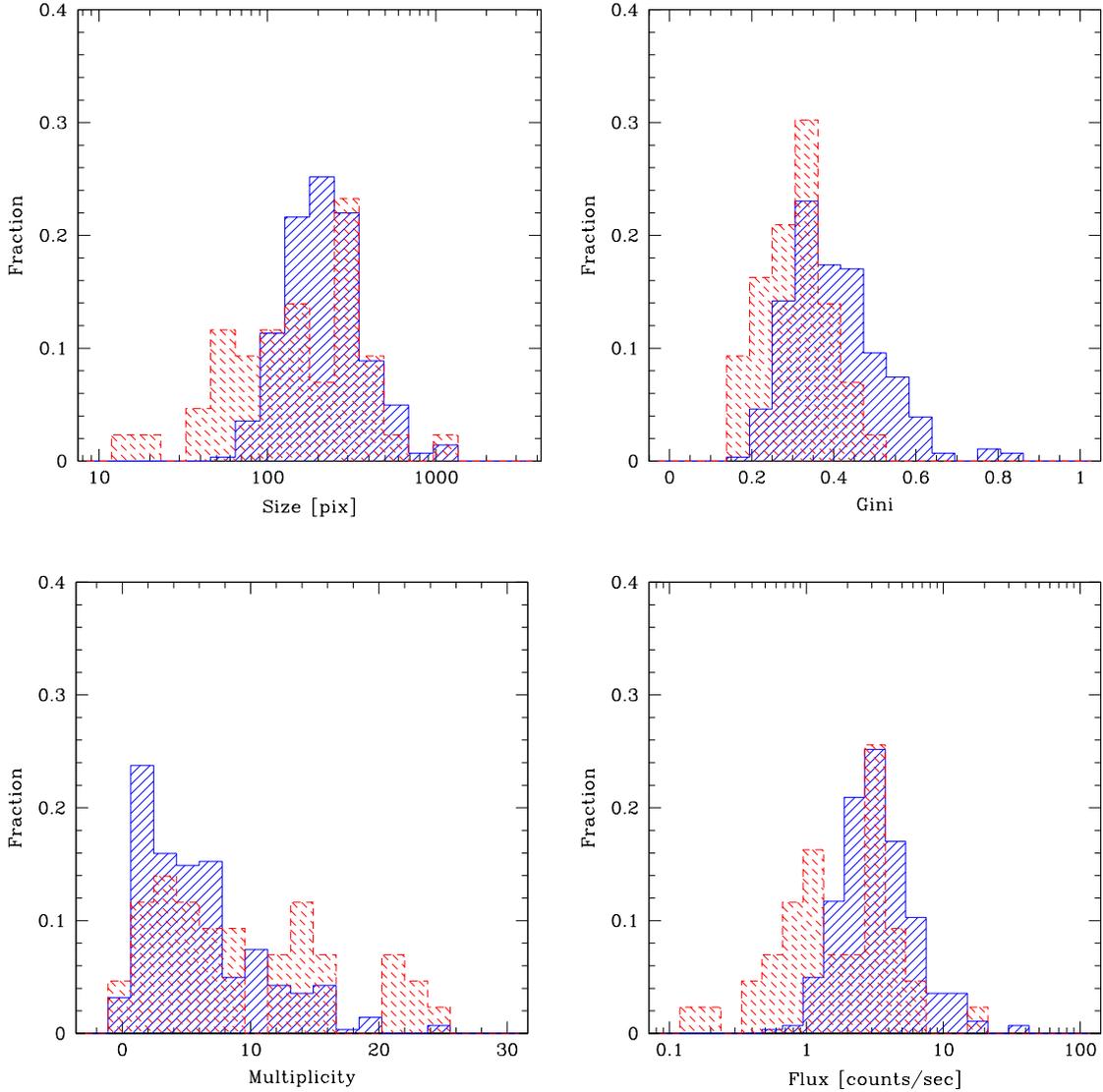}
\caption{Morphology distributions for the BX/MD sample (blue solid lines) and the DRG sample (red hatched lines).}
\label{fig:sample}
\end{figure*}

In addition to the sample of optically selected BX/MD galaxies, we have identified a sample of infrared-selected DRGs, most of which are likely to have a similar range of redshifts as the BX/MD galaxies \citep{franx2003,vandokkum2004,kriek2006}.  In this section, we will explore morphology as a function of galaxy selection method.
\newline\indent
We use two samples for our morphological comparison.  The first sample is the set of BX/MDs that lies on our four ACS pointings.  This sample includes all photometric BX/MD candidates galaxies (including those with spectroscopic confirmation), and excludes obvious foreground objects brighter than 22 mag in the F814W filter. There is a total of 282 objects in this sample.  We use this expanded sample for two reasons: it is a factor of three larger than the spectroscopic sample, and we want to control for the fact that the spectroscopic sample has a higher mean flux than the full sample of BX/MD galaxies, and may not be representative of the BX/MD sample as a whole.  The second sample consists of 43 DRGs with $K_s \leq 21$ mag.  Initial studies of DRGs selected by the criterion of $J-K_s>2.3$ (Vega) have found a mean redshift of $\approx 2.5$ \citep[e.g.,][]{vandokkum2003,fs2004}.  However, recent work indicates that a significant fraction of bright DRGs may be at $z < 2$ \citep{conselice2006,papovich2006}.  This population appears to be a mix of obscured star-forming galaxies and old, passively evolving galaxies \citep{papovich2006}.  
\newline\indent
In order to compare the two samples, we determined the mean values of the morphological parameters for each sample (Table 5), and performed a 1-dimensional KS test for each morphological parameter.  We find that, although the mean object sizes of the two samples are similar, the hypothesis that the size distributions of the BX/MD and DRG samples are drawn from the same parent population is excluded at the 99.5\% level.  This is consistent with the findings of \citet{law2007}.  As demonstrated in Figure \ref{fig:sample}, there are far more DRGs with sizes of $< 100$ pixels than BX/MD galaxies.  Both the mean values of gini, multiplicity, and total flux and the KS tests performed on these parameters indicate that the morphology distribution of the DRGs is significantly different from that of the BX/MD sample.  In particular, while 55 of the 282 BX/MD galaxies have $G > 0.5$, none of the DRGs have ginis that high.  There are more high multiplicity DRGs than BX/MDs. All of the high multiplicity DRGs have low surface brightness.  The total fluxes of the DRGs tend to be smaller than the BX/MDs.
\newline\indent
We would like to understand why the morphologies of the DRGs are so different from those of the BX/MDs.  One concern is that the DRG sample is significantly fainter, both in terms of flux and surface brightness, than the main BX/MD sample.  Of our 43 DRGs, only 17 are brighter than $\mathcal{R} = 25.5$, which is the threshold for the BX/MD sample.  Of those 17 objects, 9 are also classified as BX/MDs.  In Section 5, we showed that UV luminosity is correlated with both size and gini for the spectroscopic BX/MD sample.  If luminosity is also correlated with morphology for the photometric BX/MDs and DRGs, then differences in the morphology distributions could be attributed to the fact that the DRG sample is fainter than the BX/MD sample.  Similarly, in both samples, mean surface brightness is correlated with all morphological parameters (see Figure \ref{fig:sb} for the correlation with gini), with the possible exception of galaxy size, and so it is possible that the differences in the morphology distributions can be associated with the different surface brightness distributions of the DRGs and BX/MDs.
\newline\indent
However, we are interested in determining if there are differences in the two samples that might be due to something other than luminosity or surface brightness.  In order to control for the differences in morphology between the two samples that might be attributed to differences in surface brightness, we create two subsamples with different mean surface brightness.  The criterion in binning the galaxies by surface brightness is that the mean surface brightness of the BX/MD and DRG subsamples be approximately the same for each bin.  Therefore, we create surface brightness bins: $\langle \mathrm{SB} \rangle < 0.01$ counts/pix/sec, and $0.01 \leq \langle \mathrm{SB} \rangle < 0.02$ counts/pix/sec.  We do not create any higher surface brightness bins because there is only one DRG with $\langle \mathrm{SB} \rangle > 0.02$ counts/pix/sec.  We present the results of our statistical tests in Tables 5 \& 6.
%\newline\indent

%\clearpage

For the higher surface brightness bin, the KS tests demonstrate that the hypothesis the DRG and BX/MD morphology distributions for each parameter are drawn from the same distribution cannot be ruled out to significant confidence.  However, for the lower surface brightness bin, there are some significant differences.  Even though the mean surface brightness of both samples in this bin are virtually identical, the size and flux distributions are different at the 99.9\% level.  The DRGs are much smaller than the BX/MD galaxies.  The KS test on the gini distributions is marginal; the null hypthesis can only be ruled out at the 90\% level.  The multiplicity distributions do not appear to be different to a significant level.
\newline\indent
We would like to understand why the low surface brightness DRGs are so much smaller and fainter than the low surface brightness BX/MDs.  One possible explanation lies in our choice of observed waveband.  At $z\sim 2.3$, the F814W filter probes the galaxies in rest-frame UV, and hence, is sensitive to recent, relatively unobscured star formation activity.  If our main interest is to use morphology to determine the distribution of star-forming regions in a galaxy, the rest-frame UV may not trace the star formation of a population that consists largely of heavily obscured galaxies, such as the DRGs.  It is possible that the low surface brightness DRGs have large star formation regions, but are heavily obscured by dust.  DRGs with higher surface brightness could simply be less obscured.  Alternatively, the low surface brightness population of DRGs might represent those galaxies which have little active star formation. In either scenario, these DRGs might appear smaller and fainter in the rest-frame UV than their low surface brightness BX/MD counterparts.  With both \emph{Spitzer} observations and spectroscopic redshifts, it will be possible to perform robust population synthesis modeling for the DRGs.  Such modeling will enable estimates of star formation and extinction, and indicate the nature of these low surface brightness galaxies.
%\newline\indent

%\clearpage

\begin{deluxetable*}{lcccc}
\label{tab:ks}
\tablewidth{0pc}
\footnotesize
\tablecaption{KS test comparison of BX/MD and DRG morphologies.} 
\tablehead{
\colhead{Surface Brightness Criterion} &\colhead{Size} &\colhead{Gini} &\colhead{Multiplicity} &\colhead{Flux}
}

\startdata
all	&$0.004549$\tablenotemark{a} &$0.000021$\tablenotemark{a} &$0.013885$\tablenotemark{a} &$0.000003$\tablenotemark{a}\\
$\langle \mathrm{SB} \rangle < 0.01$ counts/pix/sec	&0.003550	&0.101108	&0.193174	&0.001701	\\
$0.01 \le \langle \mathrm{SB} \rangle < 0.02$ counts/pix/sec	&0.514717	&0.271975	&0.251298	&0.129829 
\enddata

\tablenotetext{a}{The KS probability that the morphological parameters of BX/MD and DRG samples are drawn from the same distribution.}
\end{deluxetable*}

%\clearpage

There is a possible systematic effect that might bias estimates of all of the morphological parameters.  Our original justifications for using a redshift-independent surface brightness threshold for pixel selection were that the redshift distributions of our BX/MD spectroscopic and photometric samples ought be to be similar, and that we were forced to use a redshift-independent threshold since we do not have spectroscopic redshifts for most of the DRGs.  However, if the redshift distribution of the DRGs were different from that of the BX/MDs, as perhaps indicated by recent work \citep{conselice2006,papovich2006,reddy2006}, it would be more prudent to use a redshift-dependent surface brightness threshold.  
\newline\indent
The main result from this section is that the UV morphologies of DRGs appear to be substantially different from those of the BX/MD star-forming sample. There are two main contributors to the overall differences in morphology.  The first contributor is related to sample selection.  Our two samples have different mean surface brightnesses, and morphology appears to be correlated with surface brightness for both samples.  However, even if one assembled DRG and BX/MD samples of identical mean surface brightness distributions, there would be morphological differences between the samples, notably at the low end of the surface brightness distribution.  At low surface brightness, the DRGs are substantially smaller and (necessarily) fainter than the BX/MDs.  Without more data on the DRGs, it is not possible to determine the underlying cause of these morphological differences.

\section{Discussion}
We have presented the first robust examination of galaxy
morphological parameters as a function of environment at
$z>1.5$. In particular, optical and near-IR spectroscopic
redshifts for our high-redshift galaxy candidates were a
crucial component for determining whether or not the 
galaxies in our sample were associated with the significant 
overdensity at $z=2.300 \pm 0.015$. The ability to locate
these continuum-selected galaxies in redshift space sets this work apart from
attempts to study galaxy properties as a function of
environment in high-redshift protoclusters
surrounding radio galaxies \citep[e.g.,][]{overzier2006,kajisawa2006}.
In these studies, redshift estimates for non-Ly$\alpha$-emitting 
galaxies were purely photometric, with associated uncertainties
of $\delta(z) \geq 0.5$ \citep{overzier2006}. Such uncertainties
are too large for distinguishing between genuine protocluster membership, 
and chance projection.

Previous modeling of the stellar populations of 
$1.5\leq z \leq 2.9$ UV-selected galaxies 
in the Q1700 field indicated a significantly
higher mean stellar mass and age for protocluster galaxies
at $z=2.3$, relative to galaxies in less dense environments
\citep{shapley2005a,steidel2005}. The difference
in mean age for protocluster galaxies
is consistent with theoretical expectations that galaxy-scale overdensities should collapse earlier when located within large-scale overdensities.  Along with the difference in average
stellar mass and age as a function of environment, we find {\it no} 
corresponding difference in the distribution of rest-frame UV 
morphological properties, as described by the non-parametric
gini and multiplicity coefficients, and the angular size.
We now place these results in the context
of studies of galaxy properties as a function
of environment spanning from the present epoch back to when the Universe
was roughly a third of its current age. We also highlight
promising future observational and theoretical directions.

\subsection{The Lack of a Morphology-Density Relation in the Q1700 Field}
While robust environmental trends have been demonstrated
in the nearby universe for both star-formation history
and structural parameters, we find only a connection
between environment and star-formation history in and
around the Q1700 protocluster.
%Now we return to our result that the star-formation
%histories of galaxies at $z\sim 2.3$ appear to depend on
%large-scale overdensity, while the distribution 
%of rest-frame UV morphologies does not. 
There are many factors that can account for the lack
of a strong connection between morphology and environment 
for the galaxies in our sample at $z>1.5$.
First, it is important to stress that, unlike galaxies in the
local universe, for which the structural parameters
and indicators of star-formation history (e.g., color, 4000\AA\
break, specific star-formation rate) are tightly
linked \citep{strateva2001}, UV-selected star-forming galaxies
at $z>1.5$ do not show the same connection between
structure -- at least in the rest-frame UV -- and star-formation
history. The lack of connection between morphology
and star-formation history has been demonstrated for
spectroscopically-confirmed high-redshift galaxies with modeled SEDs
in the Q1700 (this work, Section \ref{sec:phys}) and GOODS-N \citep{law2007} fields.
If galaxy morphology and star-formation history are uncorrelated at $z=2.3$, and star-formation history is the property most fundamentally connected with overdensity, then a strong environmental dependence of morphology should not be detected. 
The next significant point is that all of the morphological
analysis in the Q1700 field has been performed
on {\it HST/ACS} F814W images, which probe the
rest-frame UV for galaxies at $z>1.5$. A fairer
comparison with morphological studies at lower 
redshift, for which morphological parameters
have been estimated from rest-frame
optical images, will require imaging our full
sample in the near-IR at high spatial resolution.  Third, it has now been widely demonstrated 
that, unlike galaxies at $z\sim 0-1$,
most of which are characterized by well-defined centers and circularly
symmetric morphologies that divide clearly into
disk and bulge-dominated systems, galaxies
at $z>1.5$ generically exhibit irregular morphologies
whose physical interpretation is unclear
\citep{law2007,lotz2006,elmegreen2004b}. Since
the range of observed morphologies at $z>1.5$
is largely disjoint with that at lower-redshift,
and cannot be divided into clear subsets of disk
and spheroidal systems, at the very least caution
should be used when searching for morphological
differences as a function of environment at high redshift,
in a manner analogous to low redshift environmental
analyses.

It is also important to place the photometric properties
of our sample of high-redshift galaxies on the same
scale as used for galaxies at lower redshift. 
In the analysis of the stellar populations of galaxies
in the SDSS as a function of environment, \citet{kauffmann2004} examine several
properties such as optical color, 4000~\AA\ break strength,
and specific star-formation rate. Analysis of galaxies
in the $z\sim 1$ DEEP2 survey are presented
in terms of the average rest-frame $U-B$
color as a function of galaxy overdensity \citep{cooper2006a},
and the average fraction of ``blue'' and ``red'' galaxies as a function of
overdensity \citep{cooper2006b}
or of group vs. field environment \citep{gerke2006}. Here, ``blue'' and
``red'' refer to the two portions of the bimodal distribution of
galaxy colors, observed both locally \citep{strateva2001} and
at $z\sim 1$ \citep{cooper2006a}.

In particular,
using the set of best-fit population synthesis models for
UV-selected galaxies in the Q1700 field \citep{shapley2005a},
we compute the distributions of rest-frame $U-B$ color
and 4000~\AA\ break strength, $D_n(4000)$\footnote{This estimate
of the magnitude of the 4000~\AA\ break is used by,
e.g., \citet{kauffmann2003,kauffmann2004}, for characterizing
stellar populations in SDSS, and is defined
as the ratio of flux density at $4000-4100$\AA\ 
to that at $3850-3950$\AA.} 
for spike and non-spike galaxies, respectively.
The mean rest-frame optical color is $\langle U-B\rangle=0.67 [0.59]$
for spike [non-spike] galaxies, while the corresponding
mean value of the 4000~\AA\ break is 
$\langle D_n(4000) \rangle = 1.13 [1.20]$.
Comparison with the color-magnitude
distributions presented in \citet{cooper2006b}
indicates that both spike and non-spike galaxies
fall squarely within the color distribution
of ``blue'' galaxies in the DEEP2 sample.
The $D_n(4000)$ values for both spike and non-spike
galaxies are at the low extreme for galaxies
in the SDSS, significantly
below the transition range of $\sim 1.6-1.7$,
which divides the bimodal distribution
of galaxy stellar populations at $z\sim 0$
\citep{kauffmann2003}. Even the oldest galaxies in the Q1700
BX/MD sample, including those contained in the protocluster, 
would be classified as ``blue'' galaxies using DEEP2 and SDSS criteria.
Furthermore, the stellar population differences between spike
and non-spike galaxies are not significant when compared
with the differences between ``blue'' and ``red,'' or 
early-type and late-type galaxies, in lower redshift
surveys. These points are consequences of the young
age of the universe at $z=2.3$, the fact that 
the sample under consideration was selected in
the rest-frame UV, requiring the presence
of current star formation, and the fact that,
at $z\sim 2$, unlike in the local universe, galaxies
with stellar masses of $\sim 10^{11} M_{\odot}$ still
harbor active star-formation \citep{erb2006b,papovich2006,reddy2006}.

We must also take into account the fact
that UV-selection leads to an incomplete
census of the range of galaxy stellar populations
at $z\sim 2$ \citep{reddy2005,vandokkum2006}, missing
galaxies with red rest-frame UV colors or faint
rest-frame UV luminosities stemming from
dust extinction or an evolved stellar population.
It has been shown that the DRG criteria yields a sample
of high-redshift galaxies largely complementary to 
the set of BX/MD galaxies. While the DRG sample
appears to be composed of a mixture of
dusty star-forming, and passive galaxies \citep{papovich2006},
the typical stellar $M/L$ ratio is significantly higher
in this population than in the BX/MD sample.
Ideally, we would like to construct complete spectroscopic
samples of both BX/MD and DRG objects in the Q1700 field,
and characterize the typical stellar populations and
morphologies of combined BX/MD+DRG samples in both
spike and non-spike environments. The associated limitation 
is the difficulty of obtaining spectroscopic redshifts
for DRGs, which are significantly fainter in the rest-frame
UV than BX/MD objects of the same mass, and characterized by
lower-equivalent-width H$\alpha$ emission in the rest-frame
optical, except in the case of an AGN \citep{vandokkum2004}. 
Current and future
multi-object near-IR spectrographs on $8-10$~meter-class
telescopes will aid in the effort to obtain spectroscopic
redshifts for DRGs. Without spectroscopic redshifts,
it will not be possible to characterize the environments
of DRGs, due to the magnitude of error in typical
photometric redshifts \citep{vandokkum2006}.
By including passive galaxies in the mix, we will be
able to study the fraction of galaxies without
ongoing star formation as a function of environment,
which is more analogous to lower-redshift studies
of the relative fraction of ``blue'' and ``red'' galaxies than our previous
investigation of ``blue'' and ``slightly less blue'' objects.
%than what we've done so far with the BX/MD sample alone.
Indeed, we have only shown so far that BX/MD galaxies in the spike appear
to have started forming stars earlier than non-spike galaxies.
However, since galaxies must have ongoing star formation
to satisfy the BX/MD color criteria, we have little information
about the past or future epoch when star formation ceases,
and the related causes.

We have presented several explanations for the lack
of observed dependence of BX/MD morphology on environment,
including the fundamental differences between low- and high-redshift
galaxy morphologies, the apparent lack of connection between irregular
morphology and star-formation history for UV-selected
star-forming galaxies at $z\sim 2$, and the young stellar
populations of all the galaxies considered here when
compared with samples at lower redshift.
However, other recent observations of morphologies of
galaxies at similar look-back times suggest evolution in morphology
as a function of {\it redshift}, which may also be relevant
to the expected environmental dependence of morphology in
high-redshift protoclusters.
%would have suggested observed
%differences in the distribution of morphological
%properties for spike and non-spike objects. Indeed,
Indeed, \citet{law2007} find that the mean value of the
gini coefficient is significantly higher for
a sample of $z\sim 3$ LBGs, relative to that
for the corresponding sample of UV-selected
galaxies at $z\sim 2$, indicating more nucleated
morphologies for the higher-redshift sample.
Estimating typical sizes of star-forming galaxies from
$z\sim 1$ to $z\sim 5$, \citet{ferguson2004}
conclude that galaxies at higher redshift
are smaller, at fixed luminosity.
As a reflection of the acceleration of structure
formation within the protocluster, \citet{steidel2005} suggest that spike galaxies
attain a level of maturity at $z\sim 2.3$, that will be reached by typical 
``field'' (i.e., non-spike) galaxies at at slightly lower redshift.
Given the physical overdensity of the Q1700 protocluster, 
the morphological properties of spike galaxies should be similar
to those of non-spike galaxies at $\langle z \rangle \sim 1.8$,
as opposed to non-spike galaxies at $\langle z \rangle =2.3$
(as in the current non-spike set).  However, our small sample size,
and the relatively small offset in effective redshift ($\Delta z\sim 0.5$),
most likely prevent significant detection of 
differences in the average gini or size of spike galaxies due 
to their lower effective redshift.

\subsection{The Existence of a Color-Density Relation in the Q1700 Field}
There are multiple lines of argument to explain
the lack of difference in the distributions 
of rest-frame UV morphologies for spike and non-spike galaxies
in the Q1700 field. While there is no correlation between
environment and morphology for our sample of galaxies, 
the stellar populations of spike galaxies do appear
older and more massive than those
of galaxies in less dense environments. As discussed
above, these differences are more subtle than
the low-redshift distinction between ``red'' and
``blue'' galaxies, yet the fact remains that we
do find statistically significant environmental differences in the stellar
mass and age distributions. It is important to consider this
result along with those from the DEEP2 survey,
that the average color is redder and
fraction of red galaxies is higher in group, or denser, environments 
at $z<1.3$, but that there is 
no detectable dependence of galaxy color
on local environment at $z\geq 1.3$ \citep{cooper2006b,gerke2006}.
Given the $\sim 1$~Gyr timescale for galaxy colors to 
transform from ``blue'' to ``red'' following the cessation of star formation, 
the appearance of environmental trends at $z=1.3$ indicates that 
the preferential formation of ``red'' galaxies in denser
DEEP2 environments must commence at $z\leq 2$. For the $z=2.3$
progenitors of the galaxies probed in the DEEP2 survey, then,
no environmental trends are expected.

Our results at $z>2$ can only be reconciled
with the DEEP2 results if the 
galaxies in the $z=2.3$ protocluster do not represent
the progenitors of the bulk of the DEEP2 sample,
but rather the progenitors of galaxies
residing in richer environments by $z=1.3$.
This appears to be the case, if indeed
the $z=2.3$ protocluster virializes 
to become a $\sim 10^{15}M_{\odot}$ cluster by
$z\sim 0$ \citep{steidel2005}. We investigate
this question with the aid of the Millennium
Run simulation \citep{springel2005},
a large, high-resolution N-body simulation
of cosmic dark matter structure formation.
According to dark matter halo merger trees constructed
from the output of the Millennium simulation
\citep{springel2005,lemson2006}, the typical mass
ratio between a $\sim 10^{15} M_{\odot}$ dark
matter halo at $z\sim 0$ and its most
massive progenitor at $z=1.3$ is roughly
a factor of $\sim 6$. Therefore, by $z=1.3$,
the Q1700 spike galaxies should be residing in a 
halo of mass $\sim 2\times 10^{14} M_{\odot}$.
The DEEP2 survey is more heavily weighted towards
galaxies residing in the field or in less massive structures.
The bulk of groups featured in the DEEP2 sample of \citet{gerke2006}
range in mass from $5\times 10^{12} \leq M \leq 5\times 10^{13}$,
with a negligible fraction of systems with $M>10^{14} M_{\odot}$.
The DEEP2 sample of \cite{cooper2006a,cooper2006b} probes a 
similar range of environments. Therefore, the environment
of the likely $z=1.3$ descendant of the $z=2.3$ protocluster
is not well-represented within the DEEP2 environmental
samples, and there is no contradiction between
our observed color dependence on environment at $z=2.3$
and the lack thereof in the DEEP2 survey at $z=1.3$.

\subsection{Future Directions}
There are several promising theoretical and observational directions to pursue.  More work is required to determine the detailed three-dimensional
distribution of the protocluster galaxies, and the manner
in which they are grouped. Given that the angular 
distribution of galaxies contained within the
spike in redshift space extends over at 
least $\sim 11$ comoving Mpc on the sky, and is not
obviously segregated from non-spike galaxies \citep{steidel2005},
the protocluster defined
in redshift space does not constitute a single, well-defined,
virialized object. Therefore,
we would like to understand the mass of typical structures
containing protocluster galaxies at $z=2.3$ that will
inhabit $\sim 10^{15} M_{\odot}$ clusters
by $z\sim 0$. Based on the Millennium simulation merger trees,
it appears that the $z\sim 2$ progenitor halo mass distribution of 
$\sim 10^{15} M_{\odot}$ clusters at $z\sim 0$
typically contain between
one and five $\sim 10^{13} M_{\odot}$ dark matter halos.
Since the dark matter halo mass in which a galaxy resides
affects the amount and fate of mass and 
gas it accretes \citep{keres2005,dekel2006},
a careful understanding of the way in which protocluster galaxies populate
dark matter halos at $z\sim 2$ will shed light on their 
future star-formation histories. 
While detailed dark matter plus hydrodynamics simulations 
of massive cluster formation \citep{kravtsov2005} have been performed,
most of the close comparisons with observations have been tuned to $z\sim 0$.
We would also like to employ
a close comparison of the detailed star-formation
histories and structures of simulated protocluster and field galaxies prior
to $z=2.3$. 

Furthermore, future morphological studies of statistical samples of $z\sim 2$ in the 
rest-frame optical will provide a cleaner view of galaxy structure.  
The camera currently on {\it HST} with the minimum 
acceptable angular resolution for this
type of study is NIC2 on the Near Infrared Camera and Multi-Object Spectrometer (NICMOS),
though its small field of view 
provides a survey efficiency too low
for building up a statistical
sample of rest-frame optical morphologies.
However, the Wide Field Camera 3 (WFC3), to be
installed in {\it HST} during Servicing Mission 4,  
will enable this type of study in future years. Our detailed observations of
the environment of the Q1700 protocluster, plus a close
comparison with simulations of cluster formation tuned to $z=2.3$,
will provide a unique opportunity to learn about the seeds
of the environmental dependence of galaxy properties.

\acknowledgments
Based on observations made with the NASA/ESA Hubble Space Telescope, obtained at the Space Telescope Science Institute, which is operated by the Association of Universities for Research in Astronomy, Inc., under NASA contract NAS 5-26555. These observations are associated with program \# HST-GO-10581.07-A.  Support for program \# HST-GO-10581.07-A was provided by NASA through a grant from the Space Telescope Science Institute, which is operated by the Association of Universities for Research in Astronomy, Inc., under NASA contract NAS 5-26555.  A.E.S. acknowledges support from the David and Lucile Packard Foundation.  C.C.S. and D.R.L have been supported by grants AST 06-06912 from the US National Science Foundation and \#HST-AR-10311 from NASA through the Space 
Telescope Science Institute.  We thank Jennifer Lotz for providing us with her morphological analysis of the GOODS-N field, Mauro Giavalisco for useful discussions, and an anonymous referee, whose comments improved the paper.

%\bibliographystyle{apj}
%\bibliography{apj-jour,lbgrefs}

\end{document}